\documentclass[12pt]{article}

\usepackage{latexsym}
\usepackage{graphicx}
\usepackage[usenames,dvipsnames]{color}

\def\be{\begin{equation}}
\def\ee{\end{equation}}
\def\bea{\begin{eqnarray}}
\def\eea{\end{eqnarray}}

\def\e{\emph}

\def\nb{$N$-body problem }
\def\nbn{$N$-body problem}

\def\mb{$N$-body }

\def\case#1/#2{\textstyle\frac{#1}{#2}}

\def\id{, i.e., }

\def\s{${\mathcal S}$ }
\def\sn{${\mathcal S}$}

\def\elb{$I_\textrm{\scriptsize cm}$ }

\def\tb{three-body problem }
\def\tbn{three-body problem}

\def\isa{$V_\textrm{\scriptsize New}$ }
\def\isan{$V_\textrm{\scriptsize New}$}

\def\po{Poincar\'e }

\def\n{Newtonian }

\def\no{$N$ }
\def\non{$N$}

\def\eu{Euclidean }

\def\el{$\ell_\textrm{\scriptsize rms}$ }
\def\eln{$\ell_\textrm{\scriptsize rms}$}

\begin{document}

\begin{center}
{\Huge\bf
Gravity Without Absolutes}\large\footnote{Significantly expanded text of online talk by JB for Time in Physics International Workshop, São Paulo, Brazil, 13/05/2025, available at the link https://www.youtube.com/@PrincipiaInstituteprograms.}

\vspace{.3in}

Julian Barbour\footnote{Email: julian.barbour@physics.ox.ac.uk; Website: platonia.com} and Maria I. R. Louren\c{c}o\footnote{Email: fc56407@alunos.fc.ul.pt}

\end{center}

\abstract{By formulating his theory of universal gravitation in absolute space and time, Newton demonstrated the power of reductionism; it has been behind every major advance
in physics since then. However, the lack of significant advances over the last half-century, together with unresolved problems in the dominant paradigms of physics and cosmology, suggest that holism may need to replace reductionism as the way forward. The ontological simplicity and historical significance of Newtonian gravity makes it the ideal theory for examining this possibility. Taking Newton’s second law of motion to be the core of his theory, we show that its use within the reductionist
framework of absolute space and time introduces an infinity of arbitrary solutions if used to describe $N$-body universes rather than subsystems within them. We show
that Leibniz’s principle of sufficient reason (PSR) eliminates the infinite arbitrariness
and transforms the predicted nature of $N$-body universes; they are all shown to be unexpectedly interesting. An even greater change is enforced if one asks what principle
could determine the number of particles in the universe. The PSR suggests it
can only be infinite. We suggest that, in such a universe, correlation could replace causation while still giving the appearance that each state has been created by dynamical evolution from the past, whereas its actual explanation may lie entirely within the present state. While much work remains to develop these ideas into a viable theory, we are confident that it advances the case for holism.}

 
\section{Introduction} {\large Newton's formulation of his laws of universal gravitation in absolute space and time allowed him and subsequent researchers to study the motions of any finite number of bodies in what became known as the \nbn; it was the stimulus for the development of dynamics in all its many current forms. Despite the numerous successes of this work, it has not broken free from the reductionist framework in which it was born and challenged already by Newton's contemporaries. In particular, Leibniz argued that the positions of bodies are not defined relative to absolute space but relative to other bodies, which created what he called ``the order of coexisting things''. He also rejected Newton's time that flows ``equably without relation to anything external'' and said time is simply ``the succession of coexisting things''. Leibniz's view is clearly much closer to everyday experience but was almost completely forgotten by the great success of dynamics until revived by Mach a century and a half later \cite{Mach1893}. He made the important point that although Newton's results had been obtained by mathematics formulated in absolute space and time their confirmation was relative to the spatial framework defined by the fixed stars and a time variable created by the Earth's rotation.

Like Leibniz, Mach was a committed holist and said \cite{Mach1893}:\begin{quote}\small
Nature does not begin with elements, as we are obliged to begin with them. It is certainly fortunate for us that we can, from time to time, turn aside our eyes from the overwhelming unity of the All and allow them to rest on individual details. But we should not omit ultimately to complete and correct our views by a thorough consideration of the things which for the time being we left out of account.\end{quote}\normalsize

{\large Reacting to Ludwig Lange's proposal \cite{Lange1885} that Newton's absolute framework should be replaced by an inertial frame of reference defined locally and operationally, Mach conjectured that such a frame is actually determined by the relative positions and motions of all the bodies in the universe. This idea had a big influence on Einstein and was a major stimulus to his creation of the general theory of relativity, which he believed expressed Mach's ideas and led him to formulate what he called {\it Mach's principle} \cite{Einstein1918}. However, whether and how it was implemented in general relativity proved controversial; in fact, near the end of his life Einstein said his attempt had been misguided since ``the general type of Newton's mechanics: masses and their interaction as the original concepts $\dots$ does not fit into a consistent field theory, as will be immediately recognised'' \cite{Einstein1949}. The fact that Einstein had, for several years, worked under the belief Mach's ideas were consistent with the field theory he was developing, his `immediately recognised' is an indication of how easy it is, even for a genius, to miss the obvious, especially in the creation of a truly holistic theory of the universe. Moreover, the many different views about how Mach's ideas should be implemented \cite{BarbourPfister1995} show how difficult it is to intuit directly an adequate notion of `the All'.

Rather than attempt to do that, we have found, almost by accident, what might be a more fruitful approach: identify one after another all the defects that stem from Newton's reductionist framework in his theory of universal gravitation. We do this first under the assumption that Newton's second law of motion is correct and show that \mb initial conditions which are unquestionably Machian admit an infinity of representations in Newton's absolute framework and a corresponding breakdown of predictability. We then show how Leibniz's principle of sufficient reason (PSR) can be used repeatedly to eliminate one failure of predictability after another. The effect is like unpacking a Russian doll with each successive doll better than the one before: the PSR repeatedly allows the retention of only those solutions that belong to ever smaller zero-measure subsets of what would otherwise be deemed valid \mb solutions. Moreover, in each such step there is a clear sense in which the solutions that remain are progressively less reductionist and simultaneously more interesting. 

The greatest change in the nature of the allowed solutions comes with the elimination of the implicit and hitherto almost entirely unrecognised extrinsic scale in Newton's absolute space. In preparation for that we will conclude this introduction with definition of the key notions of scale-free physics and a first illustration of their utility. However, before that we need to say what is the final effect of judging possible \mb solutions by the PSR---none remain! We conclude that, if holistic, a universe of particles in \eu space cannot be governed by a law that is dynamical. In this paper, and in \cite{BarbourLourenco}, we argue that this is also true for a field theory. However, this does not mean that a universe governed by a rational and holistic law cannot exist and be like the one we inhabit. A hint of the kind of structure it can have comes from results associated with solutions that pass the test of containing no extrinsic scale; they will be presented at that stage of our paper.

We also note here that the first revolution in human understanding of how things are held together in the universe came with the replacement in astronomy of the reductionist Ptolemaic system, with separate eccentric epicyclic motions for each of the planets, by the holistic picture that allowed Copernicus to knit together all the relative separations and motions of the planets and to comment to the Pope that his predecessors ``could not elicit or deduce from the eccentrics the principal consideration, that is, the structure of the universe and the symmetry of its parts'' \cite{BarbourDiscovery2001}.\footnote{The use of `revolution' to characterise a major historical change stems from {\it De Revolutionibus} in the title of the book Copernicus published in 1543.}

Concluding this introduction, we now show how extrinsic scale can be eliminated in the \nbn. It is based on the quotienting first proposed in the context of dynamics in 1999 in \cite{Barbour1999} of the $3N$-dimensional Newtonian configuration space ${\mathcal N}$ of \no point particles with mass ratios $m_i$, $i=1,\dots, N$ and $\sum_im_i=1,$ in \eu space by the similarity group {\sf Sim} to obtain the $3N-7$-dimensional shape space \sn:
\be
{\mathcal S}:={\mathcal N\over{\sf Sim}}.\label{s}
\ee
Thus, \s is the analogue of ${\mathcal N}$ but for the {\it shapes} that mass points can define.

In support of the step that (\ref{s}) represents, we note here in a brief digression that the primary role of shape and effective irrelevance of scale is now recognised in many contexts unrelated to dynamics, several of them already before \cite{Barbour1999}. For example, in the late 1970s the mathematician and statistician David Kendall \cite{Kendall1984} proposed the notion of shape space mainly to analyse archaeological problems associated with the alignments of megalithic stones. Developing this work, Fred Bookstein created geometric morphometrics \cite{Bookstein1991}, which is widely used to study the shapes of historical artefacts. Similar mathematical concepts are used in facial recognition technology, which is based on ratios of separations of distinctive points on the human face; its overall size captured fortuitously in a photograph cannot be used. It is also clear that the techniques being developed in AI to enable robots to move around in the world must use analogous ratios on bodies. The recognition of the need for these techniques together with the ones in physics discussed in this paper suggests that only ratios have objective meaning. The {\it leitmotif} of this paper is ``Nothing but ratios''. It may also be noted that the shift from configurations with scale to shapes breaks down the barrier between hard science and all the arts, which, without exception, are unthinkable without the ratios that define structure and are reflected in Bookstein's morphometrics.

In response to the comments just made, the reader may well agree that the `raw data' in science are of necessity gathered through the observation of ratios but that scale comes into theory through the unavoidable use of units as in the universally agreed rules of metrology. We defer discussion of this `defence of scale' to later in the paper, where we will show that what we call `dimensionless metrology', which we base on shapes, has the potential to transform modern understanding of the second law of thermodynamics and cast doubt on the widespread belief that it is the one law of nature that will never be overthrown.

Returning now to our discussion of shapes, the simplest examples are triangles and tetrahedra; two points do not define a shape. \n solutions found in ${\mathcal N}\times{\mathcal T}$, where $\mathcal T$ is the one-dimensional space of \n absolute times, can be `projected' to \s as undirected unparametrised curves without loss of any essential information. All representations of the same physical histories in ${\mathcal N}\times{\mathcal T}$ that differ merely in the choice of coordinates and time variable are mapped to the same curve in \sn. We thereby reduce the employed ontology to the absolute minimum with a corresponding gain in conceptual clarity. From that point of view, it is important that each shape can, as it were, be taken out of the curve and put into a jumbled heap without any loss of information; the original order in \s can be recovered if the shapes change continuously. We think of each as a record of an instant of time and something real, as opposed to duration, which we see as a human artefact introduced to make the transition through the instants seem as smooth as possible and to match Newton's second law as explained in \cite{Barbour2009} (though that considered successive configurations, not shapes).

The resulting shape dynamics depends on seven fewer spatial variables, of which (as we said and will show) the elimination of scale is by far the most significant. In fact, `dynamics' is a misnomer because what remains is timeless: history is a curve in \sn, not a point moving along it at some speed in invisible absolute time. Such a `speed' is obtained using the ephemeris time that in the 1890s astronomers, doubting the reliability of the sidereal time that the Earth's rotation defines, introduced as a parameter with respect to which evolution unfolds in accordance with Newton's 2nd law (see \cite{Barbour2009}).

The most important aspect of the quotienting with respect to {\sf Sim} is the inclusion of dilatations, which do not belong to the \eu group. The \n gravitational potential \isa is not scale invariant; in fact, scale invariance is not a universal feature of modern physics and cosmology, although an interesting and little-known exception in Newtonian gravity will be discussed below. It is not difficult to see that the neglect is suspect. Consider any triangle. It will have a shape that its internal angles define and a size that can be increased or decreased without changing the angles. They are scale invariant. The same distinction exists for any number of points in \eu space. It is often forgotten, or not even realised, that within any system only angles are observable. Size presupposes an external ruler and has no objective meaning for a totality such as we take the universe to be. Ratios determine reality. That fact is especially clear on a moonless night. What we see are ratios, first, of angles between the stars, all of which are fractions of $2\pi$, the measure of great circles on the sky and, second, of brightness correlated, as astrophysics shows, with mass ratios. Reality is relational; the laws we use to describe it should be scale invariant. It is remarkable that they are not.

We conclude this introduction by outlining key properties of \s that will be important in what follows. We first show how, for $N\ge 3$, a Riemannian metric can be introduced on the $3N-6$-dimensional {\it relative configuration space} RCS, which is the quotient of
 $\mathcal N$ with respect to the Euclidean group {\sf Eucl}:
\be
\textrm{RCS}:={\mathcal N\over{\sf Eucl}}.\label{rcs}
\ee 
This is done using the intrinsic derivative introduced in \cite{BarbourBertotti1982} and later renamed by JB as {\it best matching} in 1994 \cite{Barbour1994}. This is most readily visualised in two-dimensional \eu space with just three particles with masses $m_i,~\sum_i m_i =1$; the generalisation to $N>3$ and three-dimensional \eu space is trivial and for more than three dimensions will not be needed. 

Let the three masses introduced above be at the vertices of two infinitesimally different triangles 1 and 2 and represent the first in a Cartesian coordinate system with origin at its centre of mass and position vectors of its vertices ${\bf r}_i, ~i=1,2,3.$ Now suppose the second slightly different triangle placed `on top of the first' in close proximity to it, creating thereby three infinitesimal displacement vectors $\textrm d{\bf r}_i, ~i=1,2,3.$ Now consider the trial quantity $\textrm ds$
\be
\textrm ds=\sqrt{\sum_i m_i{\textrm d\bf r}_i.{\textrm d\bf r}_i},\label{t}
\ee
which is clearly positive definite and must have a non-vanishing absolute minimum that is found by moving the second triangle relative to the first by means of \eu translations and rotations into the {\it best-matched} relative position:
\be
\textrm ds_\textrm{\scriptsize bm}^\textrm{\scriptsize RCS}={\textrm{min}\over{\sf Eucl}}\sqrt{\sum_i m_i{\textrm d\bf r}_i^\textrm{\scriptsize bm}.{\textrm d\bf r}_i^\textrm{\scriptsize bm}}.\label {dt}
\ee
This defines a metric on the RCS. As an indication of what a natural and universally applicable procedure best matching is we note that it leads to the important Fisher-Rao metric \cite{Rao1945} in statistics and also as shown in \cite{BarbourBertotti1982} to the basic structure of gauge theory. 

We now show how the same idea can be extended to \sn. For that we need to make the trial quantity (\ref{t}) scale invariant by dividing it by the root-mean-square length
\be
\ell_\textrm{\scriptsize rms}:=\sqrt{\sum_{i<j}m_im_jr_{ij}^2}\label{rms}
\ee
of the triangle we suppose fixed. We can then suppose the other triangle placed on top of it and best match the two triangles not only with respect to translations and rotations but also dilatations. We obtain
\be
\textrm ds_\textrm{\scriptsize bm}^{\mathcal S}={\textrm{min}\over{\sf Sim}}\sqrt{{\sum_i m_i{\textrm d\bf r}_i^\textrm{\scriptsize bm}.{\textrm d\bf r}_i^\textrm{\scriptsize bm}}\over \ell^2_\textrm{\scriptsize rms}}.\label {dt}
\ee
This defines a metric on \sn. While this step has been straightforward, its implication is profound because, unlike the non-compact $\mathcal N$ and RCS, \s is compact and on it the best-matching metric defines a probability measure. Later in the paper, we will show what can be deduced from this fact.

\section{Newton's `Corruption' of his Theory} We follow Poincar\'e in taking Newton's greatest contribution to physics to be his second law of motion \cite{Poincare1902}. It determines how forces change motion. However, to be applied initial positions and velocities, or an equivalent amount of information, must be supplied. This is normally done today in an inertial frame of reference that, except for being determined only up to a uniform velocity, shares the structure of Newton's absolute space and time. Such frame-dependent information can be contrasted with purely relational data of the kind Leibniz would approve. Suppose we want to determine possible histories of the simplest non-trivial example of a \n universe, which corresponds to the \tbn: three particles that in relational terms are expressed at any instant by two independently specifiable mass ratios and two internal angles of the triangle they form\id a point in the three-body shape space \s defined in Sec.~1. This is sufficient information to assign purely nominal Cartesian coordinates to the particles consistent with that shape. For convenience we can take the centre of mass of the particles, which can be deduced from the given information, to be at the origin.

Now suppose also given, for the same mass ratios, an infinitesimally different shape of the triangle. There will be an {\it irreducible difference} between the two shapes, which we call Apple and Pear to avoid implying one comes before the other. The problem with absolute time and space, which latter includes the implicit absolute scale we mentioned earlier, is that they permit specification, from the same irreducible difference, of a huge range of different initial data with which the second law can generate evolution. Assuming Galilean relativity holds, which it will if the potential does not depend on the position in absolute space, we can assume that the centres of mass coincide. Arbitrariness enters as follows:

\begin{enumerate}
\item The precedence and difference $\textrm dt$ in time of Apple or Pear;
\item Their unknown relative orientation: two for direction, one for rotation about it;
\item The unknown ratio of the centre-of-mass moments of inertia.
\end{enumerate}

Thus, even if we discount for the moment the freedom in the choice of the mass ratios, Newton's absolutes introduce five arbitrary elements if we want to transform relational initial data into \n initial data for evolution by the second law. In addition, there is an arbitrariness of a different kind that we will argue has the greatest implication of all: what principle if any determines the number of particles? We will discuss that when we have shown how to deal with the five already identified.

As regards the precedence in time in 1, we will show it is resolved when the problems with $\textrm dt$, 2, and 3 have been eliminated, so we turn to them. Whether Apple or Pear is assumed to come first, the nominal coordinates chosen for it fix the potential energy $V$. But since $\textrm dt$ is unknown, so too is the initial kinetic energy $T$ and with it the total energy $E$. That brings in a one-parameter arbitrariness in the resulting evolution. How can we eliminate it? 

For this, we turn to Newton's contemporary Leibniz, who not only advocated relational concepts of space and time but also, for related reasons, differed from Newton in being committed to the idea that we live in a rational universe. This led the two men to disagree about what might be called the ultimate reasons for things, for which both regarded God to be the arbiter. Newton believed God had complete freedom to do whatever He liked, but for Leibniz God would always have a reason for any choice He made and would choose the best option available. 

This led him to formulate what he regarded as his most important principle concerning contingent facts. He called it the principle of sufficient reason (PSR). According to it, no fact can hold, or a proposition be true, unless there is a reason for it.\footnote{Leibniz also had an analogous principle which governed logic and mathematics that he called the principle of contradiction. According to it a proposition cannot be both true and false at the same time.} Leibniz recognised that in general it would be difficult to find the reason for a particular contingent fact but the difficulty is much reduced if numbers are associated with a range of possibilities. This is for the simple reason that some numbers are distinguished; $\pi$ is one and so is $e$. However, the three that help in the problem with \mb initial data are the most special of all: 0, 1, and $\infty$.

Let us see how they solve the problem with the energy $E$. If initially we choose $E=\infty$, the system will be blown apart instantly; the special number 1 cannot be invoked in this case since the energy then has dimensions and cannot be scale-invariant; in contrast, $E=0$ is a scale-invariant fact and gives a perfectly sensible solution: only solutions with vanishing total energy are allowed.\footnote{Both 0 and $\infty$ are scale-invariant values of physical quantities measured in any chosen units.} They belong to our first zero-measure solution set. Exactly the same is true for problem 2, which concerns the unknown magnitude and direction (relative to the initial shape) of the angular momentum and means there are three further arbitrary elements; they too can be eliminated simply by requiring the angular momentum to be zero: ${\bf L=0}$. This brings us to the second zero-measure set, which is already constrained by being within the first. We mention here that, because the \n gravitational potential is invariant under the \eu group, the conditions $E=0,~ {\bf L=0}$ are conserved under the subsequent evolution. 

At this stage, familiar features of \n dynamics remain, but now within a more clearly relational setting of which we think \po would approve. Having, like Leibniz and Mach, the strong intuition that position and motion are relative, \po in \cite{Poincare1902} said that if at some instant the separations $r_{ij}$ and the rates of change $\dot r_{ij}$ of considered particles are known, one would think the subsequent evolution of the $r_{ij}$ should be determined by the laws of dynamics. However, \po pointed out that this is not so because the data assumed to be known contain no information about the angular momentum of the particles and that, as we have just noted, allows an infinity of different subsequent evolution. With evident regret, \po concluded from this that Newton did have good grounds for introducing absolute space. His analysis was important for \cite{BarbourBertotti1982} and suggested the Mach--\po principle, according to which the data envisaged by \po would lead to unique evolution by virtue of best matching. We believe that in this paper we have identified a principle even stronger than the one employed in \cite{BarbourBertotti1982}: the universe is rational and satisfies the PSR.

It is when we come to the third problem, the values of the centre of mass moment of inertia to be associated with Apple and Pear, that the application of the PSR has its greatest effect; it is destructive on the one hand but creative on the other. Although we can say that the centre-of-mass moment of inertia \elb is zero at both instants, that will yield a trivial solution since \elb will remain zero forever. However, the ratio of the \elb values at the two instants can be 1, which is both scale-invariant and satisfies the PSR. This possibility, realised by geodesic motion on shape space \sn, is explored in \cite{Barbour2003} but, as noted there, found wanting since it cannot satisfactorily reproduce expansion of the universe. It was recently pointed out to us by Pooya Farokhi that there is a shape-related reason for that. Whereas the \n configuration space $\mathcal{N}$ is non-compact, which means that \elb can become infinite (thereby allowing the universe to expand without limit in absolute space), the quotienting of $\mathcal{N}$ by {\sf Sim} includes dilatations and that makes \s compact. This, in turn, means that geodesics on \s are Poincar\'e recurrent and cannot be associated with an arrow of time, in contrast to, as we will shortly explain, the solutions we obtain after we have eliminated all \n solutions that fail the `no-absolutes' test.

Addressing that problem, we note that we have already reduced the \mb solutions otherwise regarded as allowed to the zero-measure set with both energy $E$ and angular momentum ${\bf L}$ zero. Can we reduce them even further? We can. Here again Leibniz helps because it was his (and also Maupertuis's) idea---that God would achieve his aims with the least effort---which led to the development of the principle of least action, long regarded as the foundational principle of dynamics. Among other conditions, it requires that at the limits of integration there must be a finite ratio, neither 0 nor $\infty$, of the quantity that measures the scale of the system, which is \eln, the root-mean-square length and up to a power of the total mass proportional to the square root of the centre-of-mass moment of inertia $\sqrt{I_\textrm{\scriptsize cm}}$. This poses the following question: do \mb solutions exist that do not satisfy this condition? The answer is yes. 

Karl Sundman \cite{Sundman1907} found the first in the \tb in 1907 and Block extended them to arbitrary \no a year later \cite{Block1908}. They are called total-collision solutions because in them \eln, the extrinsic \n scale implicit in the notion of absolute space,\footnote{Sundman and many other \mb researchers generally speak of space, rather than \n absolute space, but their intuition clearly matches Newton's. As is clear from the comments above, a notable exception is Poincar\'e \cite{Poincare1902, Poincare1908}.} vanishes at a certain stage of the evolution, beyond which no unique way to continue the solutions has been found and it is assumed Newton's equations break down. For a total collision to happen, two conditions must be satisfied. First, the total angular momentum must be zero (there is no restriction on the energy) and, second, the system must reach a special shape called a central configuration (often abbreviated to CC below) or relative equilibrium. Such shapes have great interest \cite{Saari2011}, both in themselves and above all because they arise from a holistic law, thereby providing a proof of principle for the kind of holism we advocate. For this reason, postponing for the moment the discussion of the total-collision solutions, in a somewhat extended digression we will say something about how, reading Leibniz's \e{Monadology} in 1977, led JB many years later, in 2011, to an interesting way to define and interpret CCs.

In his slim booklet, published in 1714 and consisting of 90 aphoristic paragraphs, Leibniz proposes a form of idealism based on the notion of monads, the name taken from the Greek \e{monas} already used by the Pythagoreans for `the One'.\footnote{The divine principle of reality, representing unity, the origin of all numbers, and the source of all things.} According to Leibniz, a monad is a simple substance, that is, something without parts though indivisible might put the idea better. Indeed, according to Leibniz the monads have structure through which they all `mirror' each other and thereby constitute the universe. The mirroring holds them together by what Leibniz called the pre-established harmony with which God brought them into existence at the creation and maintained ever after like two clocks that, once started, keep perfect time together through eternity.\footnote{Not stated in the {\it Monadology} but in earlier correspondence with Basnage de Beauval.} 

As another illustration of his universe of monads, Leibniz points out that the same town seen from different sides appears quite different and is, as it were, multiplied in perspective which, however, merely gives different views of the same town. He says that, in the case of monads, ``this is the means of obtaining as much variety as possible, but with the greatest order possible; that is to say it is the means of obtaining as much perfection as possible''. In Leibniz's holistic and entirely idealistic scheme, the views of the town introduce geometry only by analogy. However, it suggested to JB in 2011 the following way to define variety, in accordance with which it is \eu geometry rather than a divine clockwork that ensures pre-established harmony.

Consider three or more points in Euclidean space; what scale-invariant number might be chosen to characterise the extent to which they are uniformly distributed or clustered? To be scale-invariant, it must have length dimension zero. That is readily achieved by making it the ratio of two numbers that the complete set of points define, each having length dimension one. Mathematically two such numbers are distinguished: the first is the root-mean-square length $\ell_\textrm{\scriptsize rms}$,\footnote{One could argue that the mean length is simpler and should be used. However, the sum of squares that appears in Pythagoras's theorem persuades us that $\ell_\textrm{\scriptsize rms}$ is to be preferred; it certainly leads to interesting facts, as we will see.} and the second is the mean-harmonic length $\ell_\textrm{\scriptsize mhl}$. For $N$ points, let the $N(N-1)/2$ separations between all pairs of them be $a,b,c,\dots.$ With them
\be
\ell_\textrm{\scriptsize rms}=\sqrt{a^2+b^2+c^2+\dots}\label{rms}
\ee
and
\be
\ell_\textrm{\scriptsize mhl}^{-1}={1\over a}+{1\over b}+{1\over c}+\dots.\label{mhl}
\ee
Then the scale-invariant number we call the \e{variety} $V$, since Leibniz was the inspiration for finding it,\footnote{In the earlier texts \cite{BarbourKoslowskiMercati2014, BarbourEtAl2024}, $V$ is called the complexity.} is
\be
V={\ell_\textrm{\scriptsize rms}\over\ell_\textrm{\scriptsize mhl}}={\ell_\textrm{\scriptsize rms}\over a}+{\ell_\textrm{\scriptsize rms}\over b}+{\ell_\textrm{\scriptsize rms}\over c}+\dots.\label{fullbis}
\ee

The variety $V$ is manifestly a scale-invariant ratio. As such, it has an important property: for any nominal scale used to define distances and with them an \e{extrinsic} scale of the system, the large separations dominate the first factor and the short ones the second. Thus, except for a fractional residue, the variety is the number of times a shorter length fits into a longer one. This is exactly what happens in measurement. Thus, the variety measures the {\it intrinsic} size of the universe that the points define. It is intrinsic in that it is always some multiple of the set of sizes of the sub-structures within the considered set of points. No external ruler is required. This is the simple and direct consequence of scale invariance.

Moreover, the expanded form, with each term $\ell_\textrm{\scriptsize rms}$ divided by just one two-particle separation, shows that $V$ is exceptionally sensitive to a reduction in any of the two-point separations because then the corresponding term, and with it $V$, can become arbitrarily large, even infinite. This too is due to the scale invariance.

In this connection, consider the following problem: find distributions of $N$ points with, say, $N\ge 50$ (the value at which interesting effects become manifest), for which $V$ has a value that lies within a narrow preassigned range above the absolute minimum which, as a positive-definite function, it has. It is clear that it will be very easy to `overshoot' the value---too close proximity of any two points will do that. Moreover, the expanded form of (\ref{fullbis}) shows that `hitting' the target value involves the achievement of a delicate holistic balance that is established mainly between the shortest separations especially if many particles are involved. This is because $\ell_\textrm{\scriptsize rms}$, the numerator of each term in (\ref{fullbis}), is only weakly sensitive to a pair of particles getting closer to each other, but exactly the opposite is true for any of the denominators. The following general conclusion can be drawn: among the possible distributions of many points for which the variety $V$ has a given value, virtually all of them will contain many smallest two-point separations that are very nearly equal. This property will be shared by many obvious generalisations of the variety, which appears to be the simplest possible function we could choose to meet the task proposed above.

We now come to the connection with Newtonian gravity. In 2011, thanks to discussion with specialists in the \n \nbn, JB learned that, with sign reversed, the variety (\ref{fullbis}) plays an important role in it and is called the shape potential or normalised Newton potential. To see the connection with the \nbn, suppose the points are particles and, in the expressions for the factors of $V$, replace the letters $a, b, c,\dots$ by the \eu separations $r_{ij}$ to which they correspond, and add masses $m_i$ that are a fraction of a nominal unit total mass, $\sum_i m_i=1$. Then
$$
\ell_\textrm{\scriptsize rms}=\sqrt{\sum_{i<j}m_im_jr_{ij}^2},
$$
$$
\ell_\textrm{\scriptsize mhl}^{-1}=\sum_{i<j}{m_im_j\over r_{ij}},
$$
and the variety becomes
\be
V:={\ell_\textrm{\scriptsize rms}\over\ell_\textrm{\scriptsize mhl}}=-\ell_\textrm{\scriptsize rms}V_\textrm{\scriptsize New}.
\label{fullb}
\ee
The numerator in the variety (\ref{fullb}) then measures the mass-weighted size of any \n \mb system, while the denominator is minus the inverse of the Newton potential \isan. The minus sign is purely nominal; without it, \isa was, and often still is, called the force function. 

Several comments are here in order. First, everything of interest already holds for equal-mass particles. They, like particles with different masses, can be treated, as in the heuristic derivation of the variety, as points with zero size. Second, as just shown, the zero size of points, with or without mass, makes satisfaction of scale invariance trivial along with the formation of structures in the form of clusters of points. We find it hard to imagine a simpler ontology that is capable of modelling in scale-invariant terms our experience of structures around us and in the universe at large. 

We note an interesting difference between Newton's practice in his great discoveries, in which he used mass-weighted points of zero size that traced out the ellipses whose scale-invariant properties were the foundations of his proofs, and his intuition as expressed in the 1717 edition of his {\it Opticks}, in which he spoke of ``solid, massy, hard, impenetrable, moveable Particles that God formed in the beginning''. We are encouraged to use the points of Newton's mathematical principles endowed with mass ratios (and also their direct generalisation to charge ratios). We also note that trigonometry means measurement of triangles, the vertices of which are points of zero size. Beginning with its use in antiquity by Hipparchus, trigonometry was the tool by means of which humans bit by bit came to know how vast is the universe in which we live. Indeed, parallax in \eu space was the effect that enabled Copernicus to deduce ``the structure of the universe and the symmetry of its parts''. It was also trigonometry that suggested the notion of complex numbers, which have been most important in many branches of science and engineering, in particular quantum mechanics. 

We think points in Euclidean space represent the simplest ontology through which all known science can be expressed. As regards geometry, Poincar\'e argued \cite{Poincare1902} that Euclidean geometry, being the simplest that is consistent, should be retained and measured departures from it attributed to the effects of the forces that determine the solid bodies used in measurement. Einstein \cite{Einstein1921} fully endorsed his argument. In fact, \cite{LourencoEtAl2026} shows how the shortest two-particle separations that exist as a natural consequence of the holism associated with any fixed value of (\ref{fullb}) realise models of unit-length rulers. We note that Maxwell introduced field theory primarily in order not to have to explain phenomena by action at a distance, about which Newton himself had been most uncomfortable. However, according to quantum mechanics, correlations can in principle be, and experimentally have been, established instantaneously over arbitrarily large distances (though without any information being transmitted). This too suggests the possibility of a much simpler ontology without the infinity of field degrees of freedom in any finite region of space. Indeed, the fact that pixels can be used to create images of waves, both static and moving, shows that closely spaced points could produce the effect Maxwell attributed to fields.

We end our digression with two observations. The first is that the simplest way to define the variety of points in space, which Leibniz himself could have proposed, leads directly to the two functions that define the oldest non-trivial dynamical theory: the \nb of Newton's theory of universal gravitation. Simply defining how particle separations change in a scale-invariant manner by means of two functions, nothing more, nothing less, unifies the aspirations of those two bitter enemies. More relevant than this posthumous reconciliation is the fact that the special shape with which it is only possible for total-collision solutions to terminate must be one corresponding to a critical value of the variety (\ref{fullb})---the absolute minimum (the most interesting case as we will see), a local minimum, or a saddle.

The second observation is the striking connection between variety, defined purely mathematically as the prerequisite for structured experience, and gravity, the dominant force responsible for large-scale structure in the universe while the other forces create `filigree' within it. Is this a hint that the overarching principle which governs the universe is both holistic and teleological? We will have more to say about that possibility.

Concluding our digression, we now note that the condition which says the terminal shape must be critical shows that the moniker `total collision' is unfortunate since it suggests all the points really do collide at a single point, moreover with infinite speed. But if we insist only the possible shapes of Newtonian universes are real, total collision gives a misleading impression for which Newton's absolute space and time and the intuition that goes with them are responsible. It is true that $\ell_\textrm{\scriptsize rms}$ vanishes, but that reflects mathematical representation not physical reality and may be called a {\it scale fallacy}.\footnote{It applies, of course, only to observations of the entire universe by observers within it, not to relative scales of distinguishable structures within the universe.} All that observers within the universe could establish is that it is in a special shape: equilateral or collinear in the three-body problem, or one of the many, possibly infinitely many, critical shapes corresponding to central configurations (CCs) in the general \(N\)-body problem. As we said, CCs are better called \textcolor{black}{relative equilibria}.\footnote{Whether their number, for any finite $N$, is finite or infinite is Stephen Smale's sixth problem for the mathematics of the 21st century \cite{Saari2011}. With chiral triangles identified, the number is four for the three-body problem; Albouy and Kaloshin made the first progress on the problem for $N$ equal to 4 and 5 \cite{Albouy1995, AlbouyKaloshin2012}. The reader can find a good Wikipedia article on the subject.\label{smale}} \textcolor{black}{The reason is that the variety is a scale-invariant product of two functions from which, at a critical value of it, forces can be derived from each of the factors for which the (current) value of the other determines their strength and the resultant forces are always in exact balance whatever the overall size of the structure that results.}

The number of central configurations is an issue we will need to address shortly, but first we need to mention counterparts of total-collision solutions that are called parabolic-escape solutions. Jean Chazy \cite{Chazy1922} discovered them in the three-body problem in the 1920s. In them, the energy must be zero but the angular momentum can have any value. They too terminate in a shape for which the variety has a critical value. However, in the \n representation instead of all the particles colliding with infinite speed on top of each other with zero distance between any two of them, in the parabolic-escape solutions all the particles come to rest infinitely far apart.

These two kinds of solution are obviously already very special in that each belongs to a zero-measure set. But what about the solutions that, in the \n representation, are total-collision solutions with zero energy or parabolic-escape solutions with vanishing angular momentum? How will they appear when represented, without, as we said earlier, any loss of essential information, as unparametrised and undirected curves in shape space \sn? Will it be possible to tell the two apart? We do not think so. In both cases, there will just be a curve in \s that terminates at one end as the shape of a central configuration and extends forever away from it through shapes whose variety increases progressively if not strictly monotonically. These solutions, ambiguously total-collision or parabolic-escape, belong to the smallest zero-measure subset we have so far identified.\footnote{There are, of course, an infinity of curves in shape space that can be made to both begin and end at central configurations. We express doubts whether Newtonian solutions could exhibit such behaviour.} 

Besides being highly rational on the basis of the PSR test, these solutions are particularly interesting in two respects. \textcolor{black}{The first is that, by the time-reversal symmetry of Newton's laws, both can be interpreted as \n big bangs.} Their mathematical structure, if not their interpretation, was established before Hubble's observations showed that the universe is expanding (by two decades in the case of total-collision solutions). It is also worth noting that the actual state of the universe at the first instant of time is not well defined in general relativity but it and the nature of the subsequent evolution are in the \n big-bang solutions.\footnote{It is subtle because they do not have standard Cauchy-type initial conditions; see \cite{{MercatiReichert2021}}.\label{Cau}} We will return to this when considering the implications of the `no-absolutes' approach in the context of Einstein's theory. The second respect in which the considered solutions have great interest is that they solve, more satisfactorily than an entropic arrow, the problem of the existence, despite the time-reversal symmetry of the laws of motion, of the common direction of time observed throughout the universe.\footnote{There are two problems with the near universal belief that entropy increase is responsible for the arrow of time and that the second law of thermodynamics is the one law of nature that will never be overthrown. The first is that it requires the universe to have had at some time in the past an extremely low entropy for which the known laws of nature provide no explanation; the second is that an entropic arrow requires the solutions of the considered system to be able to explore only a bounded region of phase space, which is not the case with $N$-body solutions. In fact, all solutions for which the energy is non-negative have non-entropic bidirectional arrows of time either side of a `Janus point' \cite{Barbour2020}; they are significantly more pronounced if the energy and angular momentum are both zero \cite{BarbourKoslowskiMercati2014} and are expressed even more clearly in the solutions that are either of the total-collision or parabolic-escape type. One further sharpening of the arrow is still to come. As explained in \cite{Barbour2020}, the unjustified faith in universal validity of the second law stems from the failure to establish the conditions under which it holds. Gibbs did state it clearly in \cite{Gibbs1902}---the condition just given in this footnote.\label{jan}} At the same time they match well in qualitative terms the known history of the universe from its conjectured big-bang origin to the present day. 

The presence of an emergent arrow of time resolves the problem whose resolution we postponed at the start of this discussion---given the irreducible difference of two shapes from which possible \n initial data are to be constructed, which comes first in time? Within the now greatly reduced set of such initial data, the answer follows with necessity. A direction of time that corresponds to growth of structure and variety exists in \e{all} the solutions that our principles allow. On that basis we can say which of the two shapes comes first in time as revealed by actual structure as opposed to Newton's absolute time, which takes reductionism to its \e{ne plus ultra}: a line of indistinguishable points with the sole attribute of imagined direction. In the scale-invariant shape-based approach, the line devoid of distinguishing features, is replaced by shapes of ever changing structure, not entropy,\footnote{It is Poincar\'e's recurrence theorem that appears to make it impossible to reconcile a dynamical arrow of time with time-reversal symmetry of the equations of motion of many dynamical theories, including Newtonian gravity. However, a necessary condition for Poincar\'e's theorem to hold is that the solutions of the considered theory can explore only a phase space of bounded measure, which is not the case in the $N$-body problem; see the discussion in \cite{Barbour2020, Gibbs1902, BarbourKoslowskiMercati2015}.} that define direction. Shapes are the true instants of time.\footnote{Instant and instance have a common etymology.} As explained in \cite{Barbour2009}, duration---Newton's time---is an emergent magnitude whose introduction as independent variable makes the evolution through the pre-existing successive shapes match the form of Newton's second law.

It is not only the problem of time-reversal symmetry that is resolved in the approach based on shapes. At the level of objective reality, with all absolutes removed, a problem related to scale appears and is resolved. When shorn of the artificial and reductionist scaffolding of absolutes, Newton's second law of motion is, as explained above, seen to define a direction---the one in which variety increases. Will it also define a direction of increasing size, measured by \eln? We do not see how it can. If the critical shape at one end is taken to be the first shape of a \n big bang (the time-reverse of a total collision), then its size, measured by \eln, will be zero. But if it is taken to be at the end of parabolic escape, then \el will be infinite. The scale fallacy is behind both conclusions---zero and infinity simultaneously is impossible. As Newton formulated his equations (non-relationally), this shows that the only solutions which should be retained are not only time-reversal invariant but also scale-inversion invariant. 

It is particularly gratifying that when Newton's theory is represented relationally by means of shapes and all arbitrariness associated with the initial conditions is eliminated the only allowed solutions begin with a maximally but not exactly uniform shape whose intrinsic size (as defined in the paragraph after (9)) increases, with at most transient fluctuations, from then on without bound. It is remarkable that the true core of Newton's theory, stripped of all the dubious elements he added to it, predicts a universe with a history strikingly similar to our own. It must expand and have a dynamical arrow of time. We may also mention that the traditional names given to both total-collision and parabolic-escape solutions imply the opposite to what can, as in our universe, be observed within them: increase of variety in both cases.
 
We have still not exhausted what the PSR can do. We noted in footnote \ref{smale} that it is not known whether the number of central configurations, for given a \non, is finite or infinite. At the very least there are a great many, factorial already for those that are collinear. With increase of the variety chosen to define the direction of time, all these central configurations could be first instants of solutions that, in \n terms, could be either total collision or parabolic escape. Can we reduce the still vast, possibly infinite number of these solutions? Here again the PSR can be invoked if we argue that only those candidate solutions for which the variety of their first instant has the absolute minimum value are to be considered. However, even this filtering may not reduce the number of non-arbitrary solutions to just one because, except in the case of relatively small $N$, it is not known if there is a unique shape that has the minimum value of the variety. Therefore, the PSR may not be able to take us any further in our consideration of $N$-body solutions.

This concludes our discussion of how Newton’s absolutes allow his second law of motion, when applied to \(N\)-body universes, to admit vastly more solutions than it should. Before moving on to two further issues to which the PSR has bearing, however, it is worth pointing out the form taken by the equations of motion for solutions that survive the PSR filtering when represented in shape space \sn.
From the point of view of mathematical simplicity, one might expect these equations to describe geodesics, for which a point and a direction in \sn~suffice to determine the evolution. However, additional information is required. For general \(N\)-body solutions, there are the five degrees of freedom we have identified: one associated with energy, three with angular momentum, and one with \(\ell_{\mathrm{rms}}\). Most significantly, however, there is just one further degree of freedom, associated with the arrow of time and expressed through the growth of the variety \(V\).
The behaviour associated with this extra degree of freedom, which gives Newton’s theory its distinctive character as a theory of creation, is described in \cite{KoslowskiNaranjoVassallo2022}, where it is denoted by \(\kappa\). Although its effects are present in all \(N\)-body solutions except for a few with negative energy, they are manifested most clearly in the `least corrupted' solutions, which can be interpreted ambiguously as either total collision or parabolic escape. Once such solutions are initiated along one of the eigenvectors of \(V\), they all tend, despite transient fluctuations, to bend away from any geodesic currently tangent to them, with an inherent tendency to climb towards ever higher values of \(V\).
This is the arrow of time, and it bears a striking resemblance to the one we observe in the large-scale behaviour of our universe. As we shall see when discussing general relativity, there is also just one additional degree of freedom in many solutions of general relativity. But there are still two more issues to consider.

The first concerns what might be called the `quality' of the solutions. By this we mean `how long' something of interest continues to unfold in solutions and can be illustrated most simply in the (non-collinear) three-body big-bang solutions for which both the energy and angular momentum are zero. They begin with the shape of an equilateral triangle that bit by bit becomes increasingly acute as two particles separate from the third and begin to orbit their common centre of mass in a motion that approximates increasingly well Keplerian ellipses whose orbital elements define ever better an `instrument' that is a rod, clock, and compass all in one and with respect to which the third particle goes in the opposite direction with a motion that, defined by the `instrument', approximates ever better Newtonian inertial motion. Unlike the ever repeating motions of the planets, such solutions evolve---they tell a story. 

Although not expressed so cleanly, as noted in footnote \ref{jan} all $N$-body solutions with vanishing energy and angular momentum tell similar stories either side of a Janus point. In fact, there is an ‘overarching’ story of such universes within which ‘sub-plots’ develop as subsystems formed by fewer particles and separate out within the inertial frame defined by the complete system and persist for varying lengths of time \cite{BarbourKoslowskiMercati2014}. While they exist, these subsystems obey emergent local laws of motion within this frame. They can have all possible energies, momenta, and angular momenta provided they sum to vanishing totals. In Dirac's terminology, the universe satisfies constrained Hamiltonian dynamics exactly, while the subsystems, as long as they exist, approximate Hamiltonian dynamics \cite{Dirac1964}. This mathematical fact is a first step to the reconciliation of holism and reductionism.

It is worth taking a critical look at a revealing example in modern physics. Introducing in 1964 the important theory of constrained Hamiltonian dynamics just mentioned \cite{Dirac1964}, Dirac said he would start with a dynamical theory involving only a finite number of degrees of freedom, commenting that it is then only a merely formal matter to pass to the infinite number needed for a field. What he did not say is that to obtain definite results in field theory one must impose conditions at some boundary and this means that the universe on the other side of the boundary can have no detailed effect on what is on this side. This is quite unlike the situation in CCs, whose structure is uniquely determined by the properties of the points it contains (mass ratios in the examples of this paper, but charge ratios must also be expected to play a role in realistic generalisations of them). Thus, field theory is necessarily reductionistic if it is to obtain results. The $N$-body problem cannot be criticised in that way since all of its particles interact with each other as a unified totality and we have shown how subsystems emerge naturally within the totality of particles and satisfy laws that are reductionist within the complete holistic system. Despite that success, from the point of view of holism the $N$-body problem suffers from a problem related to but not exactly the same as the one in field theory. To that we now turn.

\section {Infinity} 

The entire discussion up to now has been based on the assumption that dynamical evolution governs the universe and we have modelled possible situations with a finite number of particles as in the Newtonian $N$-body problem. But what principle can select the value of $N$? Both 0 and 1 are ruled out by triviality and it seems impossible to find any finite integer that is distinguished. We are forced to say that the number of particles in the universe must be infinite. One's initial reaction is that this is not a difficulty. The $N$-body problem is well defined however large the value of $N$. The theory we want is obtained simply by going to the limit $N\rightarrow\infty$.

However, there is a problem. Newton's absolute space seems to offer an infinite stage on which the happenings of the world can unfold and the early history of the $N$-body problem confirms that view. With absolute time also assumed, absolute space is perfect for expression of the first law of motion. Is there anything simpler than uniform motion in straight line? Building on that simplicity within his absolute framework, Newton's greatest achievement was to solve the two-body problem and thereby explain Kepler's laws of planetary motion. However, Newton said that the motions of the Sun, Moon, and Earth gave him headaches and he stopped working on them.

The great development of Newtonian dynamics was taken forward by French mathematicians. In particular, in the mid 18th century Clairaut did important work on the three-body problem, for which he coined the name later generalised very naturally to the $N$-body problem. On the face of it, there seems to be no limit to what can be done. As we said, just increase $N$. But the catch is infinity. You can never get there. Hilbert's hotel with its infinitely many rooms illustrates the problem. You cannot occupy them all however long you stay; there is no $\infty$-body problem. 

But even before we address that problem, to which we will return, we need to ask whether a dynamical law of the universe modelled on the $N$-body problem, or any similar model, provides the right way to think about the universe. This has in fact been an issue in modern physics, or at least in the attempts to create a quantum theory of gravity, since DeWitt's canonical quantisation of general relativity in 1967 suggested that the quantum universe is static. There are simply unchanging probabilities for static configurations.

Two main resolutions for the clash between this result and our manifest experience of change have been proposed. The first, by Page and Wootters \cite{PageWootters1983}, invokes time-independent quantum entanglement correlations between two systems with one interpreted as the `hand of a clock' and the other the rest of the universe. There are technical issues with this proposal, above all the means by which the position of the hand of the clock is `read' in the rest of the universe. The resolution of this problem was first addressed by Wootters \cite{Wootters1984} and more recently by Kuypers and Rijavec \cite{KuypersRijavec2025}. An alternative resolution of the problem was proposed by one of us, JB, in \cite{Barbour1999} based on the idea that the Born density corresponding to the wave function of the universe is concentrated on configurations of the universe that are `time capsules', these being defined as static configurations that contain structures which can, as is the case with geological strata, be interpreted as mutually consistent records of a past. 

It is unfortunate that the idea was proposed not for shapes but for configurations, which implies the existence of an absolute scale and on the basis of what we have said that is problematic. Indeed, it was argued in \cite{Barbour1999} that the configuration space has the structure of an `inverted pyramid' with its `tip' being the point at which the root-mean-square length $\ell_\textrm{\scriptsize rms}$ vanishes and that this structure could lead to concentration of the Born density on time capsules within the `pyramid'. While mathematically at least possible, from the perspective of shape space the idea was sadly misguided; $\ell_\textrm{\scriptsize rms}$ is not defined in shape space \sn. However, \s does contain distinguished points; they are all the critical points of the variety $V$ and of any other scale-invariant generalisation of $V$. In particular, since $V$ is positive definite, all shapes that, for given $N$, realise the absolute minimum of $V$ are especially distinguished. Moreover, \s is foliated by the level surfaces of $V$ and, with its rich structure characterised above all, as we will show, by the transition from maximal but not perfect uniformity to ever greater variety, is the complete opposite of Newton's absolute space, which lacks all structure. 

Before we look at details, we consider whether mathematical structure of the kind we have just described can be the basis of a theory of what we actually experience. The biggest problem is obvious. If there is no dynamical evolution, whence comes our vivid experience of change, above all motion and our strong sense that we have a past? Our answer is based on a necessary and a sufficient condition, both of which were proposed in \cite{Barbour1999}. We have anticipated the necessity above and will develop it further in what follows. To meet sufficiency, we bring in, as in \cite{Barbour1999}, consciousness and the structure of our brains. 

As regards consciousness, it is widely accepted that conscious experience is closely associated with particular neural structures and processes in the brain. We share this general view, but wish to explore a different possibility. If time and motion are not fundamental features of reality, the experience of change might be associated not with motion itself, but with the existence of brain configurations analogous to still frames in a film. In this speculative picture, consciousness would somehow account for our experience of continuity and change, rather as a film presents a sequence of still images, although the nature of this process remains unexplained. A similar possibility might apply to other aspects of conscious experience ordinarily associated with physical processes involving motion. For example, colour vision is associated with the brain's processing of light of different wavelengths. One might ask whether the relevant physical structures could be described without treating wave propagation as fundamental. This proposal remains conjectural, and much remains to be explained about how such structures could be associated with conscious experience. We now consider a minimal realisation of this idea that may help make it more plausible.

For this we need to say more about the central configurations (CCs) that we discussed earlier.\footnote{The name was coined in German as Zentralfigur by Dziobek \cite{Dziobek1900} in 1899.} They correspond to shapes that have critical values of the variety and therefore certainly do exist in \sn. There is a substantial literature on CCs, for which we particularly recommend \cite{Saari2011}. From our perspective, however, one aspect deserves greater emphasis: their scale-invariant character. Like an equilateral triangle, a central configuration retains its shape under uniform rescaling, even though its size changes. This property distinguishes the scale-invariant description of shape from the scale-dependent quantities that enter many familiar physical laws. Despite the important role CCs have played in the development of dynamics, this distinction appears to us to have received less attention than it deserves, and we therefore discuss it briefly here.

Euler \cite{Euler1767} identified the first structures that are CCs in 1763 when he showed that if the Moon were four times farther from the Earth than it actually is then, with the Moon on the line joining the Sun and the Earth and beyond the latter, it and the Earth could revolve forever in such a syzygy with the Moon's attraction towards the other two bodies balanced by its centrifugal force. From the Earth, the Moon would be seen forever full. The James Webb Space Telescope (JWST) is currently located at the corresponding point and is yielding an immense amount of information about the universe, thanks to the scale invariance that keeps it there. Nine years after Euler's result, Lagrange \cite{Lagrange1772} established the stability of equilateral triangles with the Sun at one vertex, a major planet at another, and a test body at the third. Subsequent searches revealed Trojan asteroids both ahead of and behind Jupiter at these Lagrange points. The term Lagrange points is now used for both these locations and the position of the JSWT.

For more than three particles the number of CCs increases rapidly, in fact already factorially with $N$ in one dimension, and it is an important unresolved problem in mathematics whether the number is finite or infinite in two or three dimensions \cite{Saari2011}. More relevant from our point of view is the structure of CCs, for which the results of computer calculations for 10, 000 \cite{BattyeGibbonsSutcliffe2002} and 100, 000 \cite{TotsujiEtAl2002} particles in three dimensions were published in 2002. The aim of both studies was to determine the structure of CCs as close as possible to the absolute minimum of the variety $V$, in the vicinity of which there are a great many shapes with almost the same value of $V$.\footnote{The computational programmes are designed to find the absolute minimum, for which steepest descent programmes are used, but can easily stop when very close to it.} The obtained structures (for equal-mass particles) have a remarkably uniform distribution of the particles bounded by an almost perfectly spherical surface. There is an interesting discussion of this structure in \cite{BattyeGibbonsSutcliffe2002}, in which it is attributed to Newton's potential (shell) theorem, about which we will have more to say after we have commented on a remarkable discovery made in 2021 in connection with computer calculations in two dimensions in which CCs with variety not only at its absolute minimum but also about 1.5\% above it were also found \cite{Izquierdo2021}.

The stimulus to this came from JB's realisation, first published in \cite{Barbour2020} and not apparently recognised hitherto, that, as noted above, total-collision solutions become Newtonian big bangs under time reversal. Discussion between JB and Alain Albouy at the Paris Observatory about this early in 2020 prompted the latter to hire Izquierdo a year later to establish the form actually taken by such solutions once they leave the known initial shape. \textcolor{black}{Since, as noted in footnote \ref{Cau}, these solutions do not have standard Cauchy-type initial data, Albouy suggested that Izquierdo find two CCs, one as close as possible to the absolute minimum of the variety and one a little above.} Interpolation between the two could then be used to determine initial data whose evolution would at least be a good approximation to a Newtonian big bang.

\begin{figure}
\centering
\includegraphics[width=.9\textwidth]{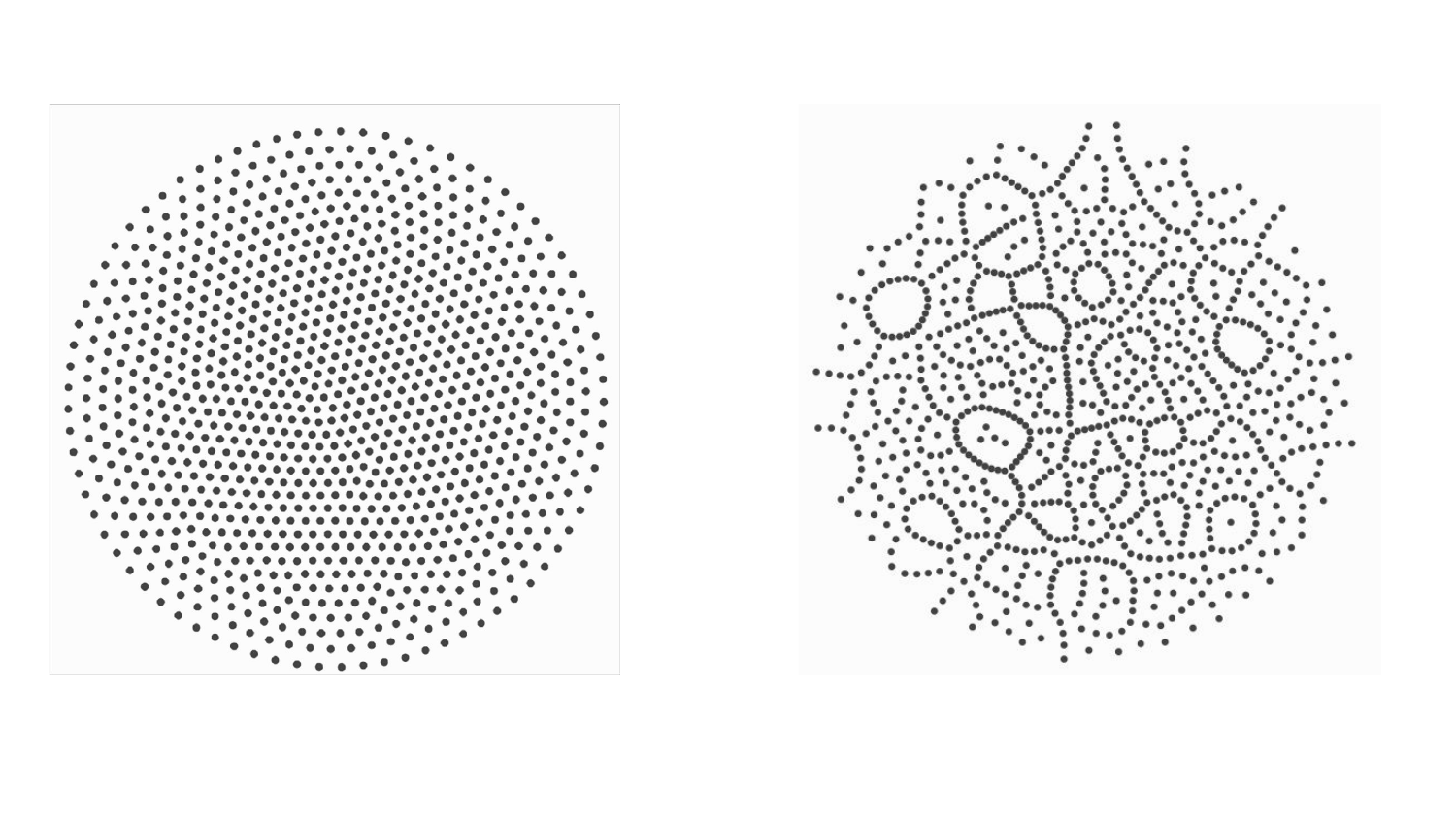}\caption{Two CCs found by M. Izquierdo. The one on the left is at or very near the minimum of the variety, the one on the right is about 1.5\% above it.}

\end{figure}

This resulted in a significant discovery in the $N$-body problem and CCs associated with it. It is illustrated in Fig.~1. On the left is a 1000 equal-mass CC in two dimensions very close to the absolute minimum of the variety $V$. Since, as we already said, there are a great many such CCs near to it (already for 100 particles let alone 1000), it is impossible to know if the minimum is realised on the left though it is certainly very near it. The discovery is the CC on the right with its remarkable profusion of filaments, many of which form closed loops with between 0 to 10 particles within them. In three dimensions with 100 equal-mass particles Izquierdo did not find any closed loops but did typically find a filament of six or seven particles. 

Having learned about this work and JB's interests, ML looked for two- and three-dimensional CCs in larger systems of particles \cite{LourencoEtAl2026, LourencoBarbourLobo2026}. Besides extending Izquierdo's CC results in two dimensions to 5000 particles, for which very similar structure was obtained (shown in \cite{LourencoBarbourLobo2026}), ML also found two CCs in three dimensions for 1000 particles, one very close to the absolute minimum of $V$ and the other at 1.5\% above it; we will give details about it below.
 
These results show that the critical points of at least one scale-invariant function correspond to interestingly structured shapes in \sn. We now present evidence that they model geological strata and perhaps even the neuronal structure in our brains. For this, based on discussions between Tim Koslowski and JB during the last two or three years, we define the age $a(s)$ of any shape $s$ of $N$ particles with given mass ratios as the excess of its variety $V(s)$ over its absolute minimum $V_0$ for the same particles:
\begin{equation}
a(s) = V(s) - V_0.\label{a}
\end{equation}
In Fig.~1, the left-hand shape is at or very near $V_0$, while the right-hand shape has $V$ approximately 1.5\% greater; accordingly, their ages are 0 and 0.015, respectively. These numbers are conceptually very different from conventional measures such as the number of orbits of the Earth around the Sun. Ages expressed in terms of ``13.8 billion years since the big-bang'' have no meaning within this geometric definition. In fact, metrology always has a conventional basis, since humans choose its units, which are not pure numbers like the age (\ref{a}) but have dimensions and limited use (the caesium atoms that currently define the second did not exist until well into the history of the universe and the first supernova explosions and neutron-star mergers).

\begin{figure}
\centering
\includegraphics[width=1.01\textwidth]{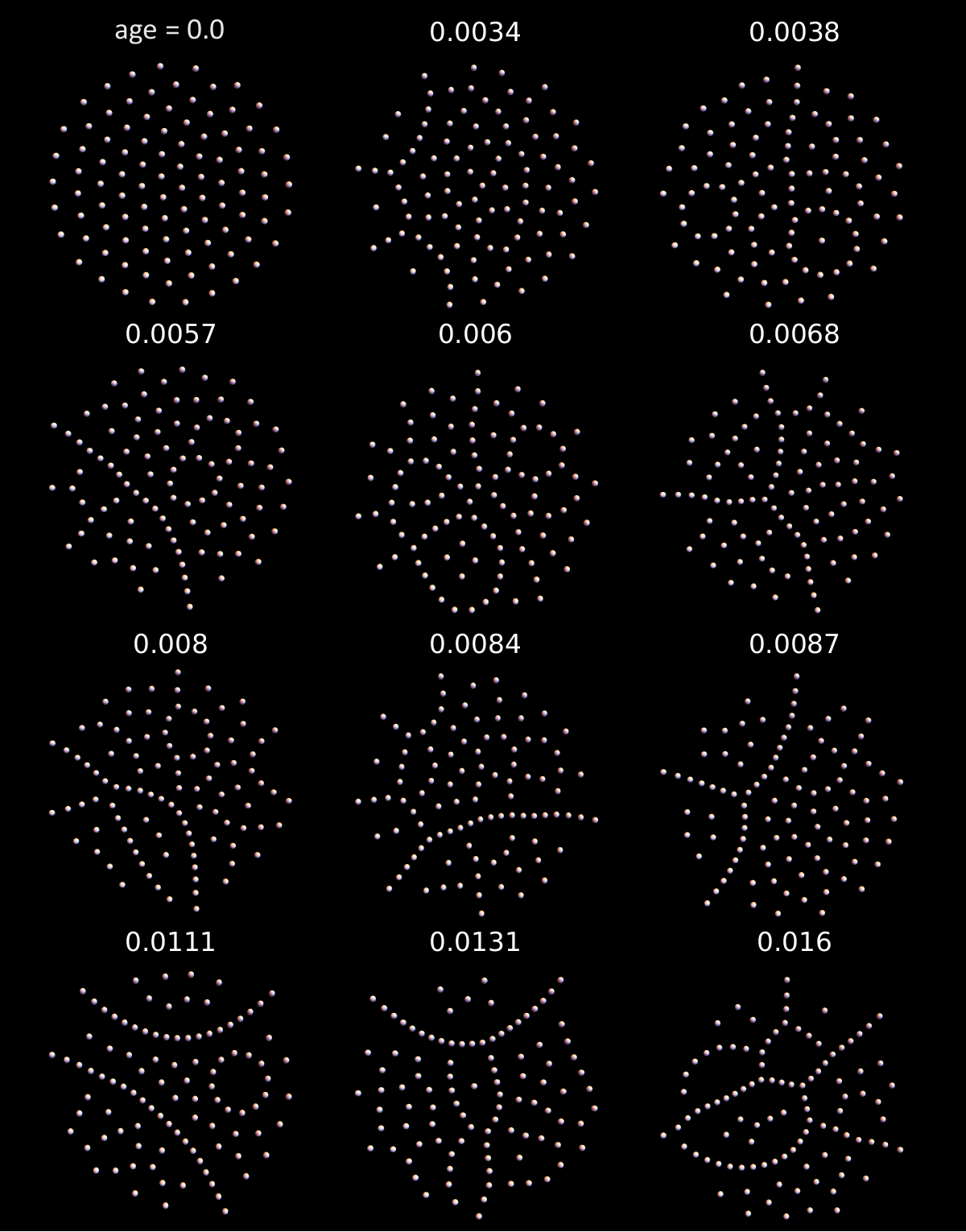}
\caption{Twelve central configurations of 100 equal-mass particles at different ages. We thank Hemant Shukla for this figure.}
\end{figure}

It is particularly interesting to see how the structures of shapes change with age---they grow. We demonstrate the effect in two dimensions, in which it can be directly seen, and then explain how it is somewhat corrected but less directly visible in the world's actual three. In the twelve equal-mass 100-particle CCs of Fig.~2, the reader will see that the first obvious structures emerge at age 0.0034 from near perfect uniformity at age 0 and thereafter become ever more complex. Moreover, as anticipated above, all but the first can be interpreted as time capsules as defined in \cite{Barbour1999}: static structures that, like 
 strata, appear to contain mutually consistent records of a past. Thus, the structure at age 0.0038 could be partially `fossilised scissors' that had been discarded due to damage which broke off one blade and bent the other. At age 0.0057, the conjoined circles suggest the fossilisation has gone further; the blade has completely decomposed. Meanwhile, the filaments can be interpreted as live vegetation at various stages of growth. 

What makes geology a real science is its ability to obtain mutually consistent estimates of the ages of different structures that exist in the present age. In all cases this is based on the structure of objects. We demonstrate the possibility of doing this with at least partial success with the CCs of Fig.~2 and then discuss the improved situation in three dimensions. The basis for success in both cases is the ability to identify distinct substructures within a complete CC and to determine the ages of both the substructures and the complete CC, whose age is necessarily the greatest.
For any unambiguous substructure, the mathematical basis of our theory will permit its age to be calculated from ratios of the separations between the points in it. The obtained age will of necessity be less than the age of the complete CC and be greater or less than that of other substructures. However, examination of Fig.~2 suggests that this seemingly simple theory cannot be applied in general because few substructures can be unambiguously identified. For example, in the CC with age 0.0057, are the conjoined circles a single substructure or two that can only be distinguished with ambiguity? In only a small fraction of cases, mostly those of filaments, are problems absent.

Although much work is needed to confirm our conjecture, we believe the two dimensionality of the CCs in Fig.~2, which makes structure manifest and suggested the idea we are proposing, simultaneously makes disentangling it into elements difficult to the eye and equally so with a computer. The limited results that we so far possess indicate that at least at 1.5\% above minimal variety the only structure, apart from single points, in the so far studied 1000-particle CC in three dimensions consists of filaments of various numbers of particles as listed there. Thus, there is no problem with the identification of individual substructures and determination of their ages. We likened above the filaments in two dimensions to growing vegetation; their branching structure does suggest that. The filaments in three dimensions might be called `worms'. 

Under the assumption that our theory is on the right lines, the central task that we see is identification of the appropriate functional form of the variety $V$; the one so far used is more or less the simplest conceivable. Whereas local gauge theory was the guiding principle that led to the Standard Model of particle physics, ours must be the finding of a variety function that gives rise to all known structure on all scales of the universe at its successive ages as defined by (9). It must therefore reproduce the known properties of the universe since the big bang. We believe only one symmetry will be needed: three-dimensional scale invariance. The work will be done by that and the variety function. Its choice will amount to selecting its `ingredients'. Those of the shapes so far presented were the simplest: equal-mass points.

To show how easy it is to create diverse shapes, the ingredients for the look alike of Saturn in Fig.~3, long interpreted as a CC in which the gravitational attraction with which Saturn acts on the small bodies in the rings is balanced by the centrifugal force that acts on them, were one heavy point mass for Saturn, seven much smaller masses for the `shepherd' moons known, in the dynamical picture, to create and stabilise the gaps between the rings, and a great many very light masses for the ice particles in them. For the least-descent calculation to the CC from a random distribution of the ingredients, all points had zero size, not those of the image. Also, the added colour reflects how we think about consciousness: qualia supervene on mathematical structure that ratios define. Even sizes supervene. 

\begin{figure}[h!]
\centering 
\includegraphics[width=0.6\textwidth]{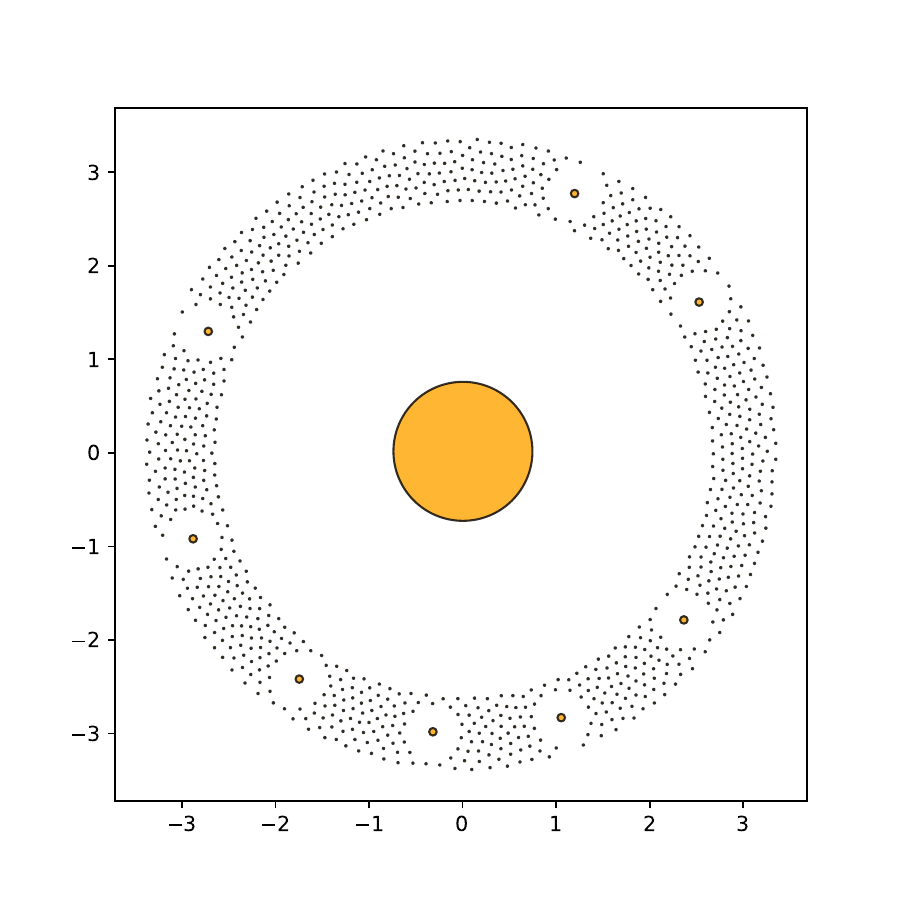}
\caption{Central-configuration image of Saturn with `shepherds' in the rings.}
\end{figure}

However, we must emphasise that the image in Fig.~3 cannot be said to be a true representation. Saturn is not a single point of very large mass but a volume containing a huge number of individual particles with masses typical of nuclei and the same is true of the shepherds and ice particles. To have any hope of an `honest' creation of Saturn and, indeed, all macroscopic bodies, the definition of the ingredients of what might be called the world function must surely involve at least one large ratio but the evidence from gravity indicates that cannot be between masses. However, there are good candidates. The first is the great difference in strength between gravitational forces, on the one hand, and those generated by electric charges and nuclear forces, on the other. There is also the fact that neutrino masses, which according to current theory play an important role in large-scale structure formation, are, while as yet unknown, tiny---less than a millionth of the electron mass. There are therefore at least two very large ratios; the standard model of particle physics also provides more, if not so large. Their use in our theory matches our ontology; as in conventional physics, we accept mass and charge ratios but not time and dynamics. Correlation, not causation, is to play the explanatory role.

The proton--electron mass ratio is large but not huge and a first step to `honest' shape creation of known structures would be to exploit it in an attempt to determine the known shapes of molecules. On that we have as yet made only limited progress, mainly because we have not yet found a variety function that reflects electromagnetic effects, without which determining actual molecule shapes is impossible. However, we have had some success in creating two-dimensional `gravitational molecules'. For this purpose, we defined six 100-particle variety functions, in each of which fifty particles had unit mass and the remaining fifty had masses 2, 80, 160, 420, 1200, or 2000 times as large, the last value being chosen to approximate the electron--proton mass ratio. We show in Fig.~4 the resulting shapes for  mass ratios. Except for the first, the light particles do cluster into the form often shown of molecules. Among these, the configuration with a mass ratio of $1/80$ is particularly striking and, visually, the most interesting. This example illustrates how changing the mass ratios can give rise to remarkably different structures, even in such a simple gravitational system. Moreover, we believe it shows, as we suggested earlier, that a theory based on shapes can break down the barrier between the arts and the hard sciences. 

\begin{figure}[htbp]
    \centering

    \includegraphics[width=0.44\textwidth]{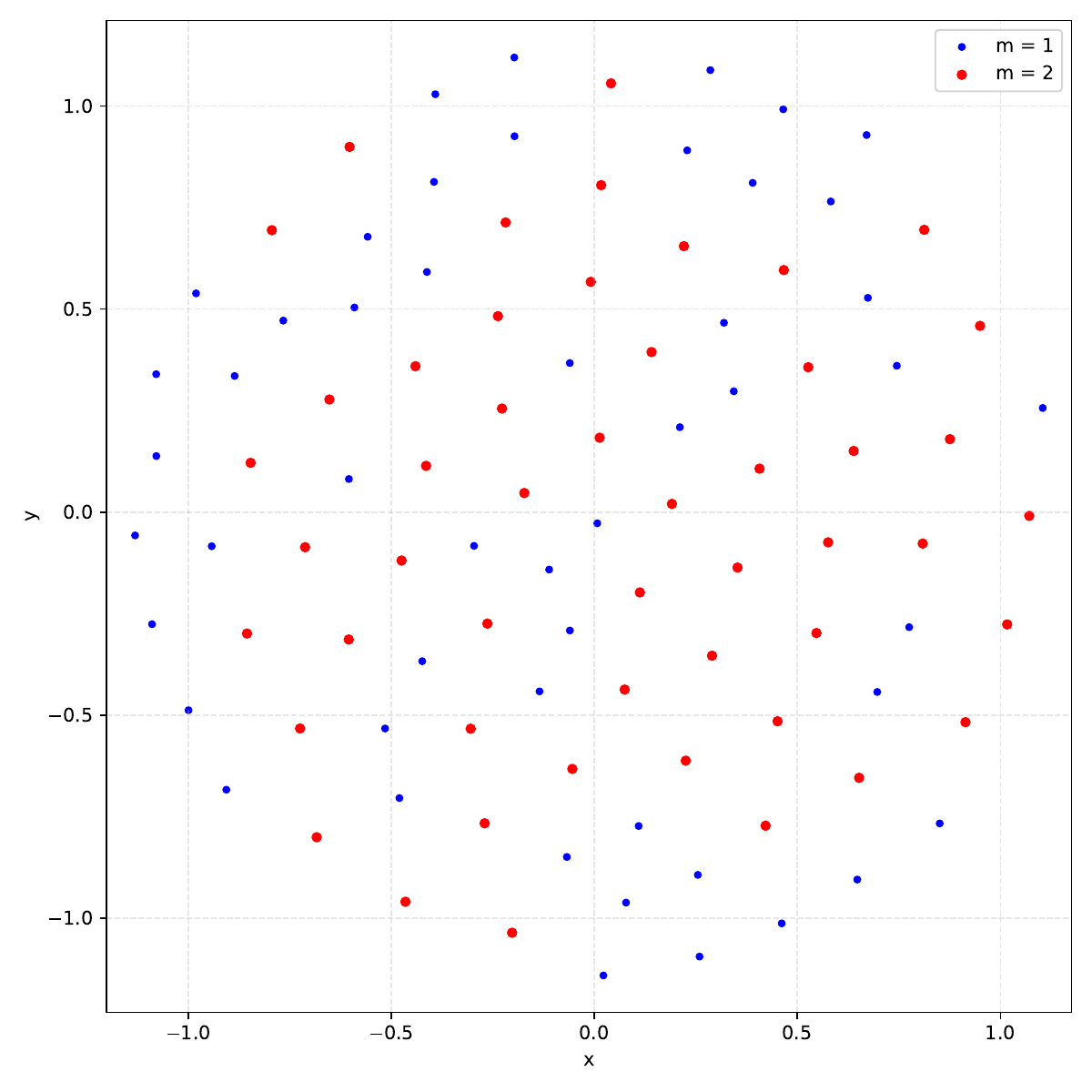}
    \hfill
    \includegraphics[width=0.44\textwidth]{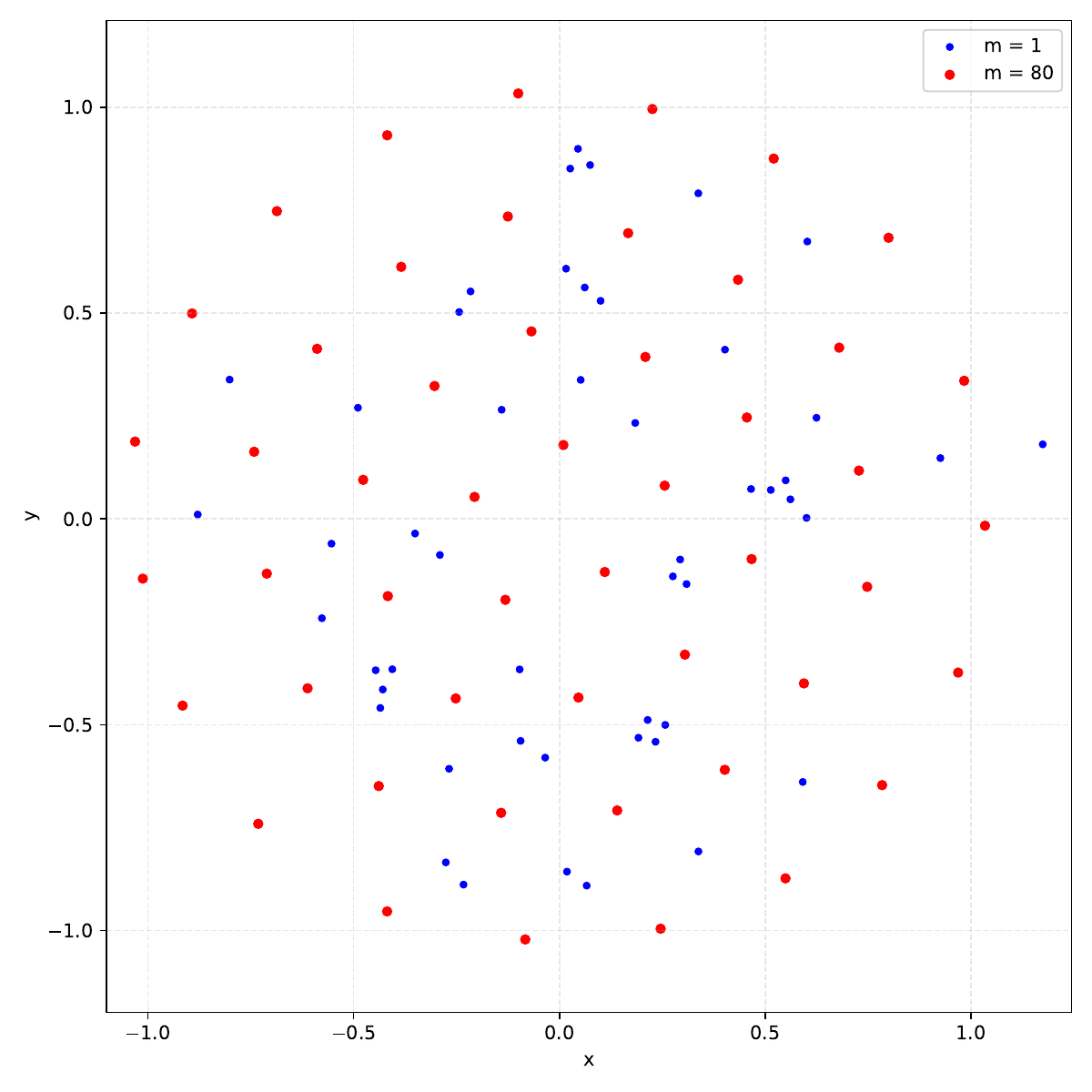}

    \vspace{0.2cm}

    \includegraphics[width=0.44\textwidth]{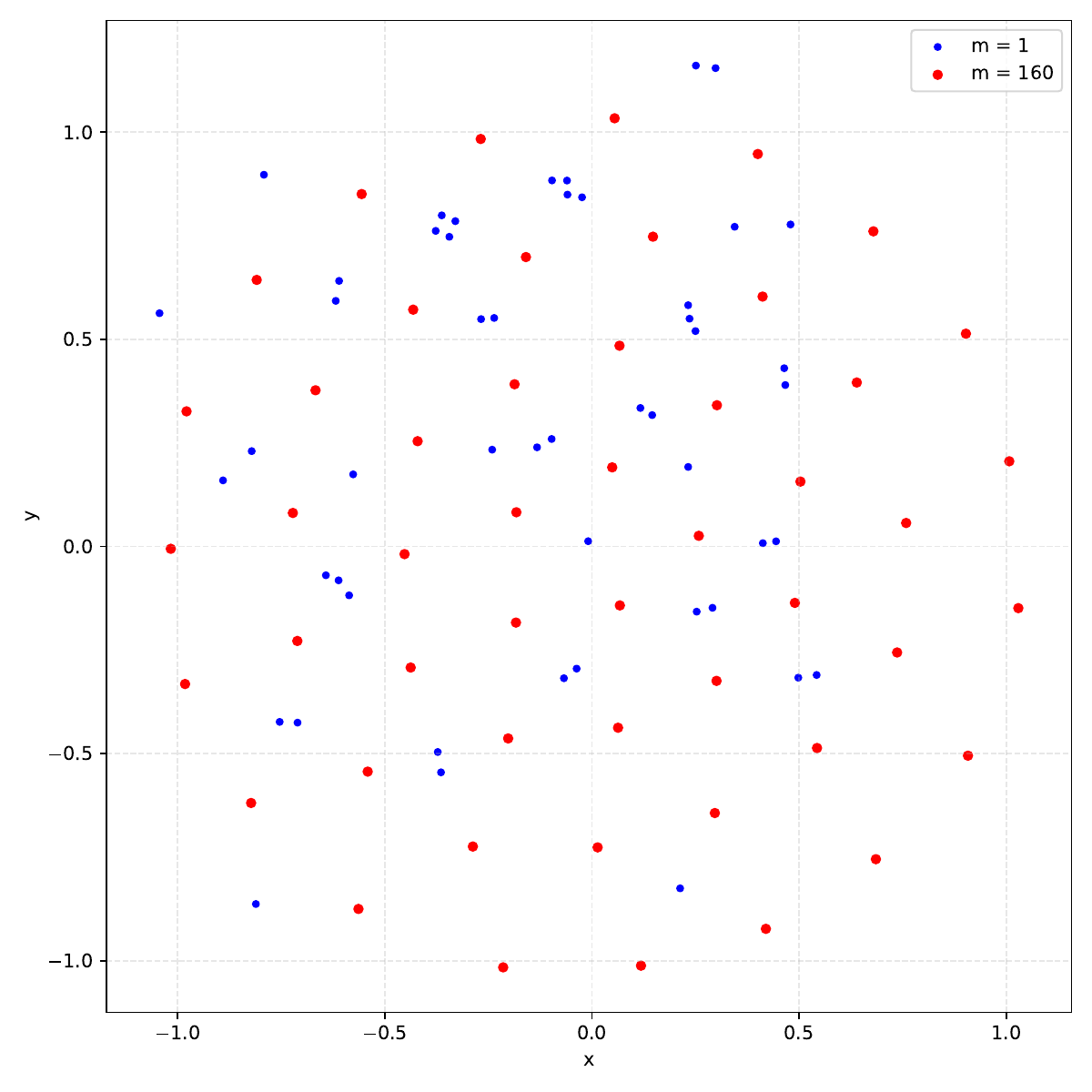}
    \hfill
    \includegraphics[width=0.44\textwidth]{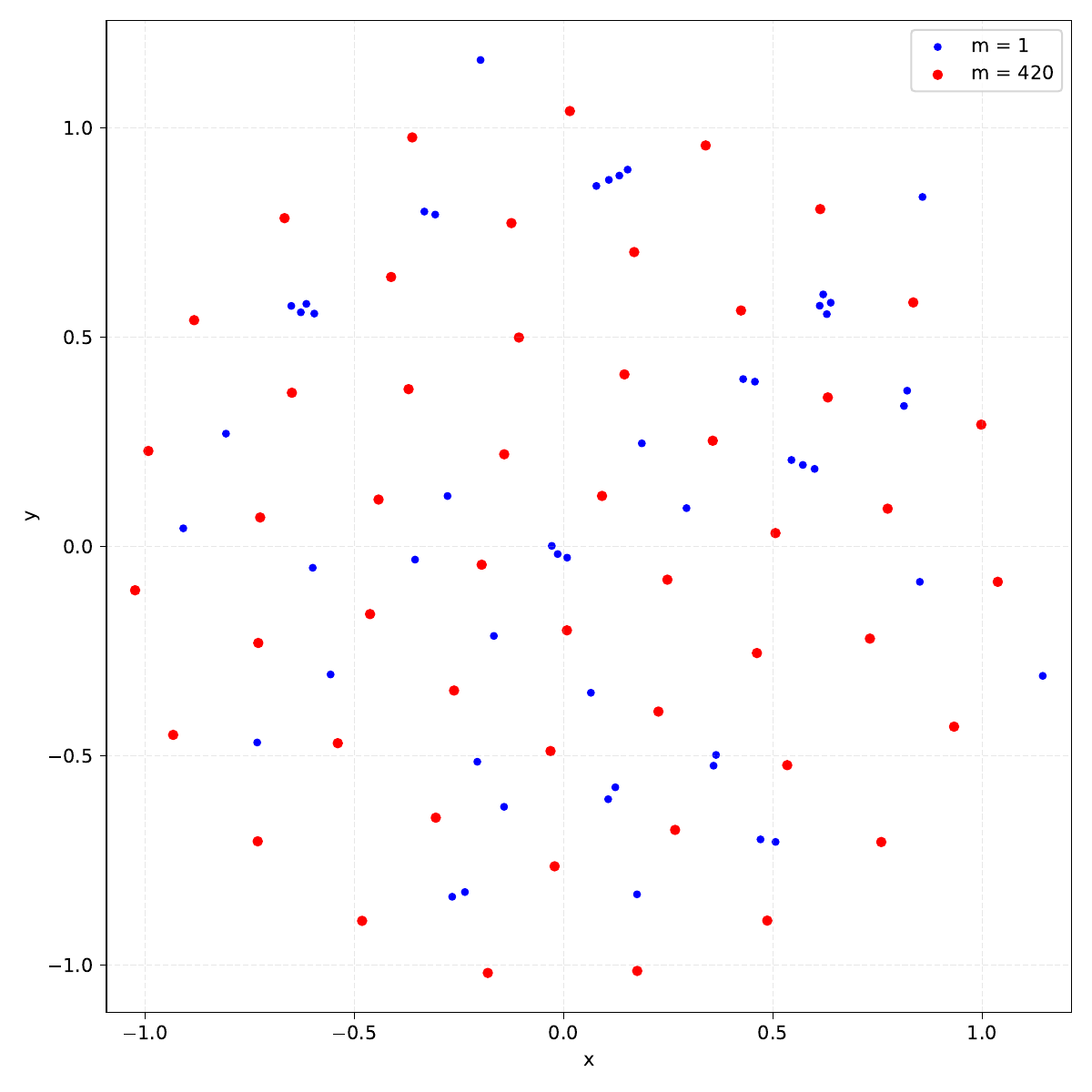}

    \vspace{0.2cm}

    \includegraphics[width=0.44\textwidth]{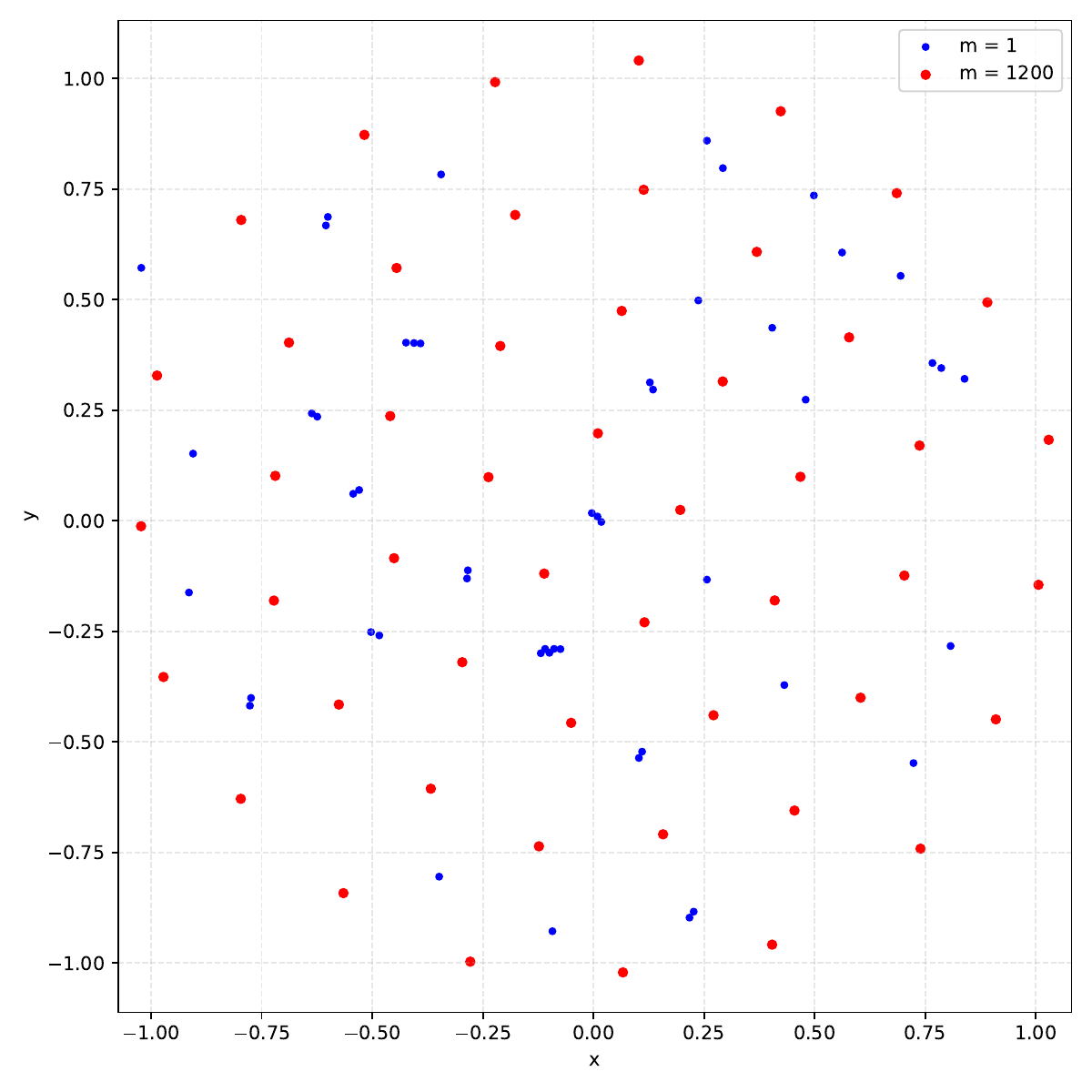}
    \hfill
    \includegraphics[width=0.44\textwidth]{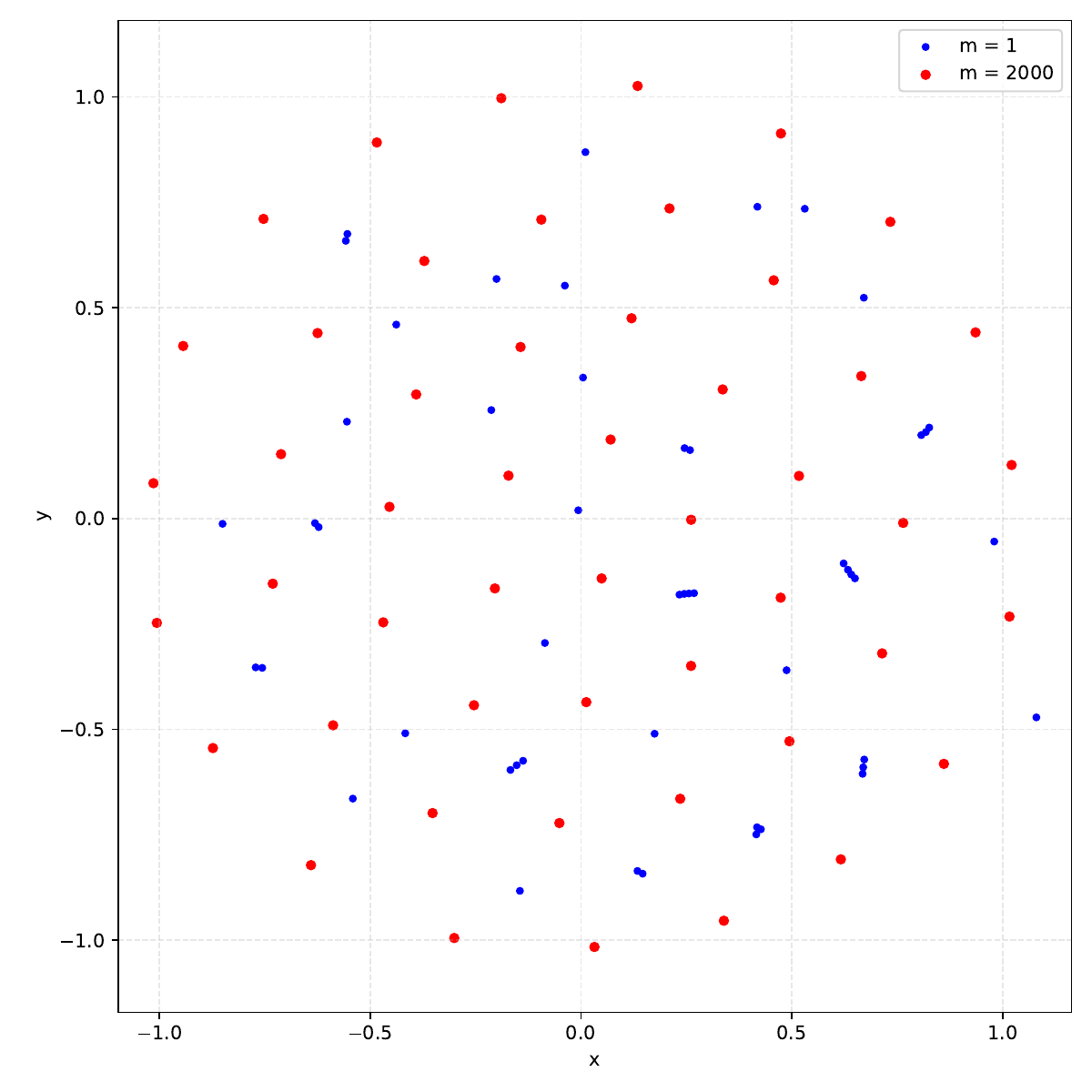}
    \caption{`Gravitational molecules' with different mass ratios of the `atoms'.}
\end{figure}

We believe the material we already possess gives hints of nothing less than a first-principles theory of creation. We noted above that as the age of the structures in the CCs increases as they grow. The effect discovered by Izquierdo and illustrated in Fig.~1 prompted JB a few years ago to check the etymology of creation; it has nothing to do with \textit{creatio ex nihilo} but comes from the Proto-Indo-Germanic verb {\it ker}: to grow. The difference between its nature as required in two and three dimensions we are discussing may be revealing. If something is to grow, indeed is manifested to do so in our scale-invariant theory, it needs space and directions in which to do so. We think there being only two for the CCs is responsible for the frequent branching that is present in Fig.~2 while being absent in the results so far obtained in three dimensions. A factor that is relevant in both cases is the space that is available; it is pre-determined by the theory and is not a space quantifiable in the familiar sense by an implicit absolute scale. What counts is the intrinsic size. We will consider the implications of this fact after the interim summary that we now give.

Taken together, the 12 CCs match qualitatively in their overall form and progressive structural growth what we know about our universe at its different ages since the big bang. However, there is a big difference between the current causal and dynamical explanation given by physicists and cosmologists for that and the origin of the images in Fig.~2. Each was determined independently of the others; the appearance of evolution in time is illusory. Increase of structure found at shapes with critical values (minima and saddles) of the variety, nothing more and nothing less, is the ordering principle in Fig.~2. We think it possible that Newton's absolutes have led, through making reductionism the dominant paradigm in physics, to an entirely misleading picture of reality and the nature of the universe. So far as we know, nothing comparable to the creation in Fig.~2 of what looks like the outcome of causal evolution exists in the literature (apart from early stages of our programme). Of course, much detailed work will be needed if the facts and ideas we have presented in this paper are to be developed into a serious holistic alternative to the current reductionist paradigm of modern physics and cosmology. However, we do believe we have established a certain plausibility of the ideas. 

Taking them further, in our companion paper \cite{BarbourLourenco} we question whether wave functions exist and note here some supporting evidence for that suggestion. For example, in quantum theory in the position representation, particle positions are distributed around a mean in accordance with a probabilistic law. At least qualitatively CCs exhibit an analogous effect. Because one corresponds to a critical value of the variety, which is necessarily special, so is the corresponding shape. However, shapes with the same particles and slightly different values of the variety will, for that reason, be almost as special. This was confirmed by accident by Hemant Shukla while looking for CCs with a programme that for some reason stopped just before reaching the prescribed proximity to a CC, about which it is worth saying something.

In a CC, described in standard \n terms, the gravitational force exerted on each particle by all the others is directed exactly towards the centre of mass of the complete system with a strength that increases linearly with the distance. If the particles in a CC are initially at rest, they will then fall homothetically (without changing shape) under their mutual gravity until they all collide at the centre of mass with a vanishing value of \eln. This fact can be used to check whether a distribution of particles found numerically is to good accuracy a CC; one simply sees if they do fall from rest in the obtained structure as they should. If it has been accurately determined, deviation from the required homothetic behaviour sets in only just before the total collision at the centre of mass. In the case of imperfect determination, the particle trajectories very soon become tangled like spaghetti.

This was how Shukla found several $N$-particle shapes that were only near but not at a CC. He found several structures that failed the test just described. As an indication of the remarkable shapes that do exist near critical values of the variety we show the one in Fig.~5. It suggests a tentative alternative explanation for what are regarded as established facts in quantum theory: probabilities in the outcomes of position measurements following wave-function collapse. Simply because a CC has a critical value of the variety, there must be near-critical shapes close to it in the shape space \s of the considered particles. As we pointed out in Sec.~1, there is a well-defined probability measure on the shape space of any given set of particles. The possibility discussed in our companion paper \cite{BarbourLourenco} that effects attributed to quantum theory might be explicable without wave functions relies among other things on this probability measure.

\begin{figure}[h!]
\centering
\includegraphics[width=.5\textwidth]{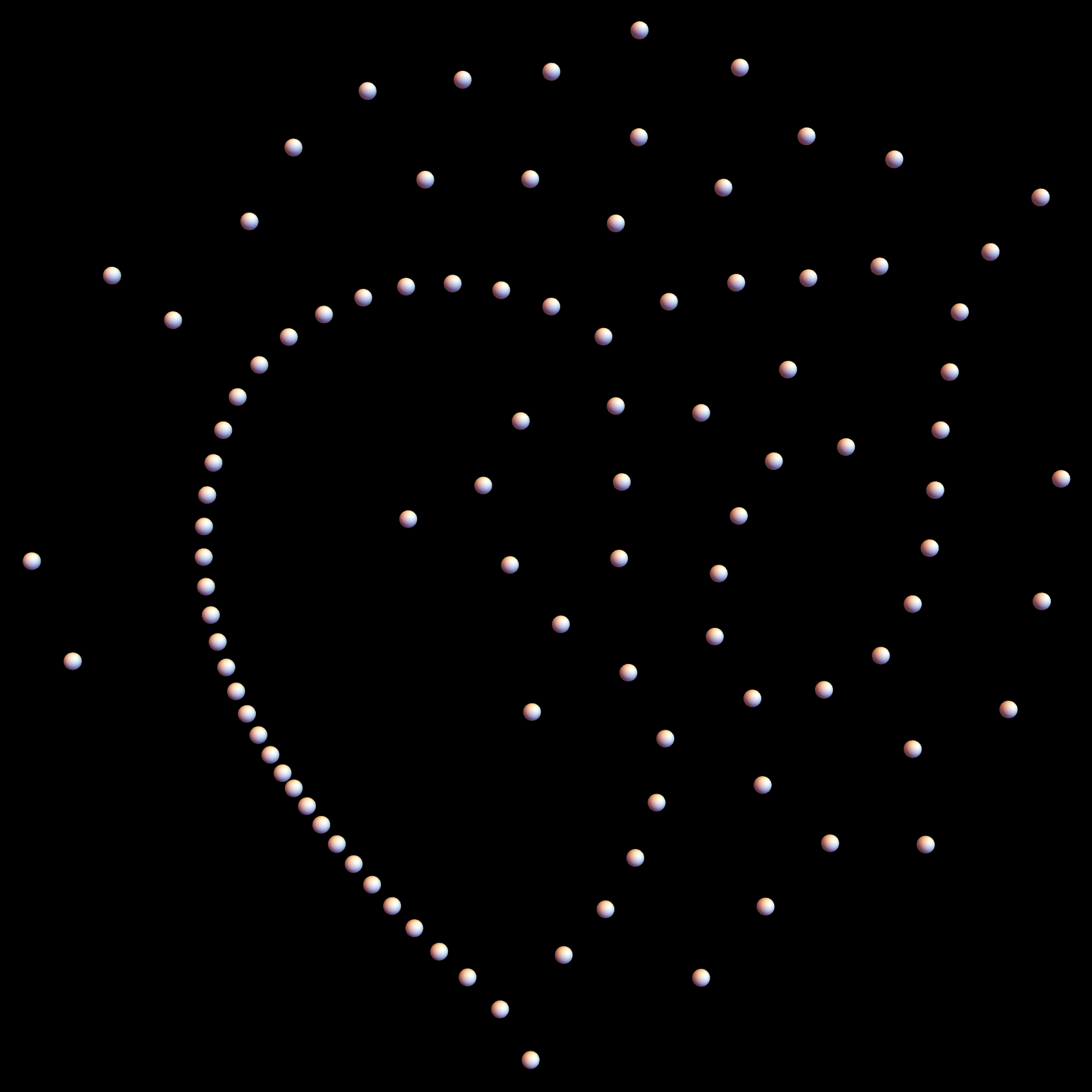}\caption{The shape near but not at a central configuration found by Hemant Shukla and called by him The Broken Heart.}
\end{figure}

\begin{figure}
\centering
\includegraphics[width=.8\textwidth]{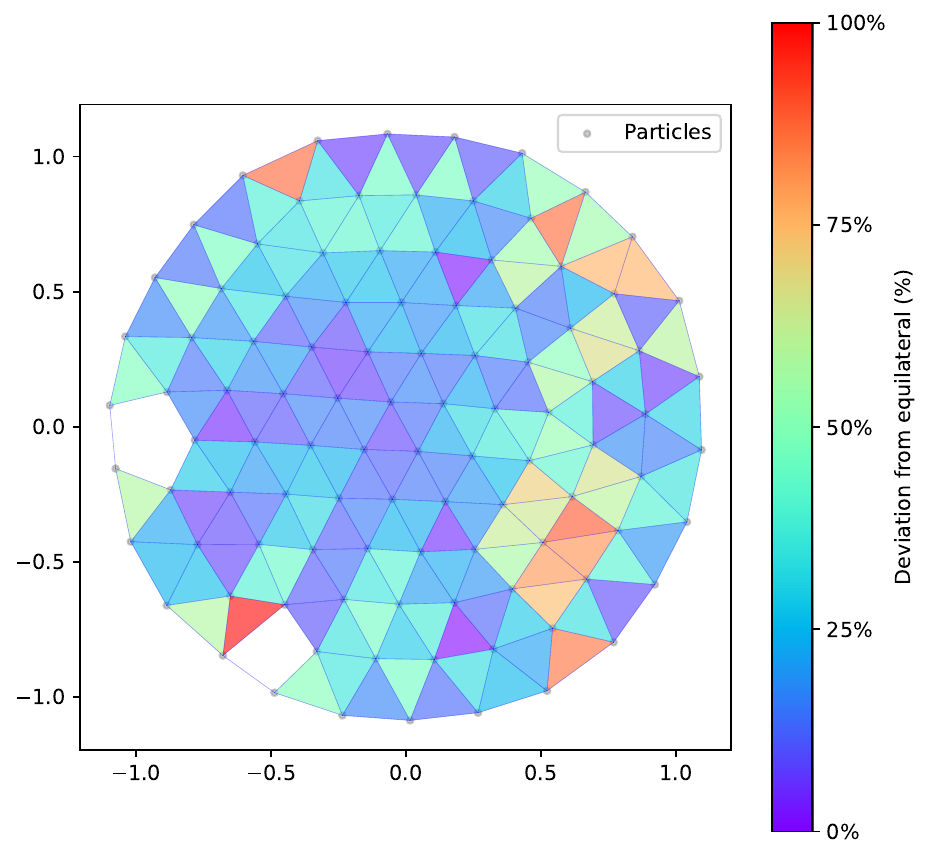}\caption{A two-dimensional CC with age at or very nearly 0 decomposed into triangles colour-coded by the amount they are not equilateral.}

\end{figure}

We conclude this part of the discussion by noting what, from a reductionist perspective, looks like a generalisation of the theory of close packing. As is well known, this arose from the problem posed to the mathematician Thomas Herriot around 1587 by Sir Walter Raleigh (a pirate as well as a knight) of how to stack identical spherical cannon balls in such a way that they occupy the least possible volume (in 1611 Kepler advanced a famous conjecture about the solution recently confirmed by a computer proof). The solution of such problems plays an important role in crystallography and the theory of Bernal liquids. In those contexts, both the shape and size of the objects to be stacked are specified; the problem is to determine the least possible volume into which they can be packed. The theory of central configurations generalises such problems to include determination of not only the shapes and relative sizes of the entities but also the volume into which they are to be packed. To illustrate this, we use, as throughout this paper, what we believe is the simplest conceivable ontology with which the form of the universe can be explained: particles of zero size weighted with mass (and charge) ratios.

Figure 6 shows the solution to the problem of finding a CC with minimal variety formed in two dimensions by 100 equal-mass particles. The resulting distribution appears to admit interpretation as the solution to a generalised close-packing problem in which the objects to be packed are triangles with nearly equilateral shapes, all with slightly different internal angles both above and below 60$^{\circ}$, the determination of which is part of the problem (in three dimensions the shapes will approximate regular tetrahedra).\footnote{The triangulation was obtained by defining two particles as neighbours whenever their separation was no greater than a prescribed radius. A graph was then constructed, with particles as vertices and links connecting neighbouring pairs. Triangles were identified as sets of three mutually connected vertices. Each triangle was coloured according to the standard deviation of its three side lengths: triangles with more nearly equal sides have smaller deviations, while more distorted triangles have larger deviations. The colour scale was normalised to the range from 0 to 1, from the smallest to the largest deviation. White regions are areas not covered by any of the identified triangles, because the neighbourhood criterion does not necessarily produce a triangle at every location.} However, this is a reductionist interpretation that seeks an explanation for individual shapes whereas the true holistic interpretation is that the intricately interconnected shape of the whole universe is determined. Moreover, the appearance of size is illusory. Nothing objective is changed by magnification or diminution of the image in Fig.~6. That leaves the actual truth, the intrinsic scale unchanged. 

We use the opportunity that Fig.~6 gives us to return to our earlier comment that, in contrast to the situation in general relativity, the nature of the initial shape of \n big-bangs is uniquely determined and contains the seeds needed for subsequent growth of macroscopic structures whereas in GR they have to be created in a rather ill-defined way by inflation and two-parameter fitting. We present the visualization of the \n seeds as an indication of how holism could trump reductionism. With three-dimensional scale invariance taken as fundamental symmetry, we believe the shape of the universe at all ages will emerge in much the same way as in Fig.~6, though of course in three dimensions.

We now return to the important issue we mentioned earlier but did not then address of how it might be possible in our approach to create a theory consistent with our principles in which the number of particles in the universe is infinite. That required the prior discussion of CCs and the notion of age, which are the basis of our theory of structure creation. It will also require something that remains to be settled in definitive form---a theory of light, or rather an explanation without light of the phenomena that we attribute to light. Heuristic principles that, we believe, should lead to it are in our companion paper \cite{BarbourLourenco}, but we point out here that we never actually see light, only things whose visible presence we attribute to illumination by light. In \cite{BarbourLourenco}, we therefore replace this account in terms of light with one based on correlation, which we assume to be between things in the universe at large and structures in our brains with which our conscious experience is in psychophysical parallelism.

After that preliminary comment, we now describe at least one way in which our theory developed for finitely many particles can be extended to the infinitely many needed for a satisfactory cosmological theory. As a first step for that, we add to our definition of age, which is well suited for considerations related to a finite number of particles, the notion of {\it specific age} as a concept better suited to considerations in which there is no upper bound to the number of particles. The aim of the new notion is to reflect the fact that it is not the number $N$ of particles that determines the value of the variety but the values of the $M=N(N-1)/2-1$ relative separations between them (weighted as necessary by the mass ratios and charge ratios of the particles). We therefore define the specific age $\bar a$:
\be
\bar a:={a\over M}.\label{sa}
\ee

Modified as necessary by our timeless and shape-based approach, we base our theory of an infinite universe on the condition adopted almost a century ago by cosmologists in the form of the cosmological principle (CP): in sufficiently large regions, all with the same size, the structure of the universe should `look alike' everywhere. At each point in the universe, its appearance should be homogeneous and isotropic. In fact, a major issue in cosmology today is that the universe as currently observed, at least within the Hubble radius, is not isotropic; there is a clear dipole anomaly. Our approach offers at least some hope of resolving the problem and involves two different things. 

Figure 1 illustrates the first. If one supposes a `dime' with a diameter of say 1/10 of the CC on the left placed anywhere except over its edges and discounts the radial increase of the particle density, it will clearly cover areas that look alike everywhere. For the CC with greater variety on the right, the `dime' will need to have a proportionately greater diameter, perhaps 1/3 of the CC. Regions covered by a smaller `dime' will clearly not look alike. \textcolor{black}{Moreover, in the CC on the right of Fig.~1 there are many regions with `sub-dime' diameter within which a distinguished direction---a dipole---can be identified.} If the kind of theory we are proposing is on the right lines, this suggests the current observations extend over a region too small to confirm the CP. 

However, such observations can still support the theory and the form of the CP it predicts because the foundational assumption of the theory is that it creates variety, the form of which will depend not only on the functional form of the variety but also the age. What we see on the right in Fig.~1 is `nuggets' of variety all different in detail but creating a uniform appearance on a large enough scale.
If the theory allows observations to be made on that scale, direct verification of the CP will be possible. This is where our comment about never seeing light but only things we attribute to its effect becomes relevant.

\textcolor{black}{In cosmology the Hubble radius defines the boundary of the observable universe and corresponds to infinite red shifting of spectral lines. The evidence for the dipole anomaly is of necessity deduced at finite red shifts. This means that the CP can still hold in regions that are observationally inaccessible. Moreover, indirect evidence that it does can still be found in observations at finite red shifts, provided that the theory predicts not only direct confirmation in regions corresponding to infinite red shifts for terrestrial observers, but also precise predictions of structures observed at finite red shifts, especially as the red shift approaches infinity.}

The reader might with reason think this is a pious hope but everything in science depends on evidence and that cannot be obtained without material instruments. The ones relevant for astronomy in general and cosmology in particular typically have the same arrangement: in the first place a telescope, then prisms to split wavelengths, next emulsions on which spectral lines are formed, after which their position-dependent intensity is measured, and finally a computer printout is examined by the eye of an astronomer whose retina is an integral part of their brain.\footnote{When we see the colours of a rainbow in the sky, water droplets in the sky and the retinas in our eyes play the role of all the material elements in this account.} That they experience anything has to be attributed to psychophysical parallelism. It is theory, not direct experience, which persuades astronomers that propagation of light at finite speed links all these stages. The actual facts are ultimately relationships between the various aspects of what astronomers experience through sight and their theory, which of necessity is always provisional.

We do not deny that the evidence for propagation of light at a universal speed is very strong. However, one of the most remarkable experimental developments in recent years has been confirmation of the theoretical prediction, in principle already made in the 1930s, that quantum entanglement correlations can be established instantaneously over arbitrarily large distances though without transmission of information. This suggests the theory of light and its effects should be based on correlation, not dynamical causation associated with propagation in time. The difficulty with this idea is the undeniable time lag between the accepted instantaneous creation of correlations and the appearance of the evidence it has happened. What this requires in our framework based on the notion of shapes with age is creation of something like Einsteinian space-time through the `horizontal stacking' of shapes of different ages in which the shape of an observer in one age reflects global structure of earlier ages.

With the variety (\ref{fullb}) so far considered, it creates in the two dimensions we are here considering characteristic structures within regions much smaller than the sampling `dimes'. Much the most characteristic thing is the existence of filaments with many nearly equal distances between successive points. In two dimensions these close up into loops, but, in the numerical studies so far completed, filaments also exist in three dimensions that are not loops but wind about each other as shown in Fig.~8. These are firm verifiable facts that emerge repeatedly and reliably from numerical studies; thus the kind of theory that we have in mind could satisfy the CP on a large enough scale but not be confirmed by telescopic observations if for some reason the theory treats light in such a way that observations cannot be made of structures at which the CP does hold.

In the current LambdaCDM framework this is achieved as a perturbation of the completely homogeneous, and hence featureless, FLRW cosmology that also became standard nearly a century ago. Data taken from observations of the CMB are superimposed on the homogeneous FLRW background and evolved forward in time as space expands. Since space, unlike things, is invisible, we find this problematic and possibly subject to what we have called the scale fallacy since the most `visible' evidence for expansion of the universe is that the average distances between the galaxies is increasing compared with their average diameters taken as `rulers'. In fact, intrinsic as opposed to extrinsic expansion of the universe with scale-invariant underpinning is clearly seen in Fig.~2. If in each central configuration the shortest two-particle separations are taken to be rulers, then measured by them it will be readily seen by eye that the overall size of the CCs increases with age. That, we think, is the true nature of the universe's expansion.

How can we extend this effect, the origin of which derives from our `golden rule' that nothing but ratios may appear in theory, to an infinity of particles? Our proposal was stimulated by the explanation mentioned earlier of the remarkable uniformity of the CCs found in the numerical study of \cite{BattyeGibbonsSutcliffe2002} (illustrated in Fig.~7) and attributed there to Newton's potential (or shell) theorem. According to it, the force exerted by a spherically symmetric mass distribution on a point mass outside it, is the same as if the entire mass were concentrated at the centre of the distribution, while the mass in a shell outside a certain distance from the centre exerts no force anywhere within the shell. 

\begin{figure}[h!]
\centering
\includegraphics[width=.9\textwidth]{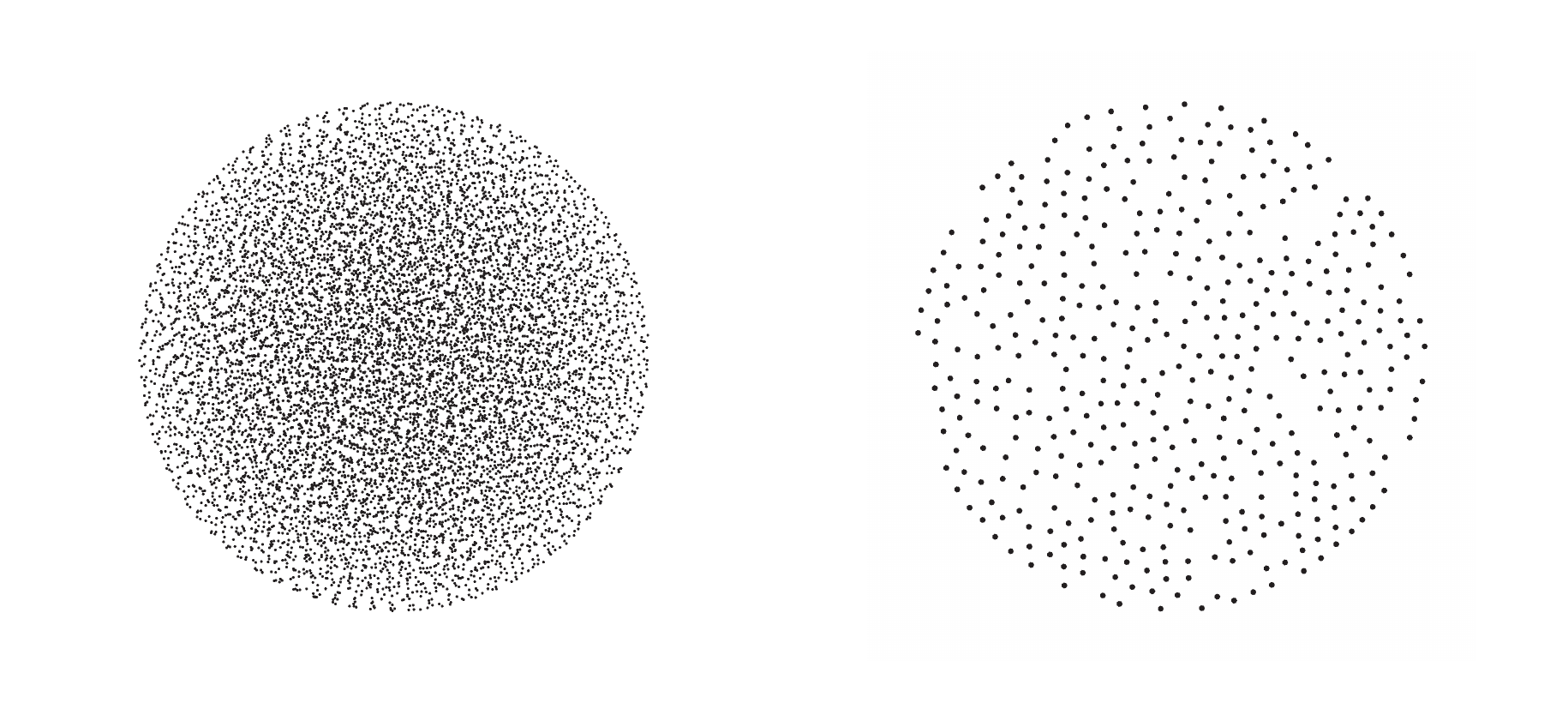}\caption{A central configuration of 5000 particles (with equatorial slice through it on the right) found numerically and kindly made available by Richard Battye.}
\end{figure}

While the explanation given in \cite{BattyeGibbonsSutcliffe2002} for the uniformity of the particle distribution in Fig.~7 was the stimulus for the ideas we are now proposing, we express things somewhat differently. First, we are not considering radially uniform solids, but distributions of points. Second, CCs do not owe their existence to Newtonian $1/r^2$ forces alone but, as pointed out earlier, to a balance between them and repulsive Hooke forces whose strength increases linearly with distance, with the two forces always being in exact balance whatever the overall scale because they are derived from the scale-invariant variety at one of its critical points. In fact, in what is to come, we propose the use of the variety as a {\it sampling function} to explore, at least in principle, the degree of correlation in regions within a distribution of infinitely many points in infinite space.

We note first that the $N(N-1)/2-1$ relative distances between $N$ points in Euclidean space are necessarily correlated since they are determined by $3N-6$ Cartesian coordinates. Also the mere fact that we defined the variety $V$ as a measure of the extent to which considered particles are clustered as opposed to uniformly distributed means that $V$ will reflect the different ways in which the correlation is manifested---through clustering or uniformity. Moreover, in an infinite space in which infinitely many particles are distributed we can always sample regions bounded by a surface that is topologically spherical and contains the same number $N\ge 4$ of particles (4 since a tetrahedron encloses a volume in three-dimensional space while a triangle does not). Let us call such a region a `balloon'. Wherever such a balloon, for chosen $N$, is placed, the variety and age of the particles that it contains can be calculated. Since the variety of \no particles with the same {\it ingredients}, by which we mean the proportions of particles with given mass ratios (and charge ratios in extensions of the theory),\footnote{As regards our proposal that the nature and properties of the particles should be part of the law that governs the universe, we note that the predictions of the current Standard Model of particle physics are confirmed to great accuracy but the particle species of the model are not determined by the theory and have to be taken from observations. We speak of ingredients since the creation of a CC does have a culinary parallel: one first creates a random distribution of specified proportions of particles with different properties which is transformed numerically by steepest descent into a CC. Moreover, the great number of similar cakes that result from baking parallels the vast number of CCs associated with the same inputs.} is a measure of their correlation, it is to be expected that the contents of the great bulk of such balloons will `look alike' if the values of their variety are nearly equal. To develop a proper theory, we need to define `look alike' rigorously. 

Evidence that we so far possess allows us to show how our principles enable us to do this for shapes in three-dimensional \eu space for the two lowest ages so far explored numerically. At age 0, the results confirm that the CCs can be decomposed in many different but qualitatively equivalent ways into structures that are nearly regular tetrahedra. As explained above, the analogous decomposition into nearly equilateral triangles in two dimensions is illustrated in Fig.~6 for a CC with age at or very nearly zero.

Numerical results in three dimensions for a 1000-particle CC found by ML at variety about 1.5\% above the absolute minimum and illustrated in Fig.~8 provide clearer evidence in support of our approach. It contains just two kinds of structure: regions filled with more or less uniformly distributed points and filaments containing variable numbers of points that clearly stand out because they seem to have `created a tunnel' through the unstructured points. \textcolor{black}{The filaments are coloured according to the inverse of their mean nearest-neighbour separation, with different colours representing different values of this quantity.} The remaining particles are shown in grey.

\begin{figure}[h!]
\centering
\includegraphics[width=0.6\textwidth]{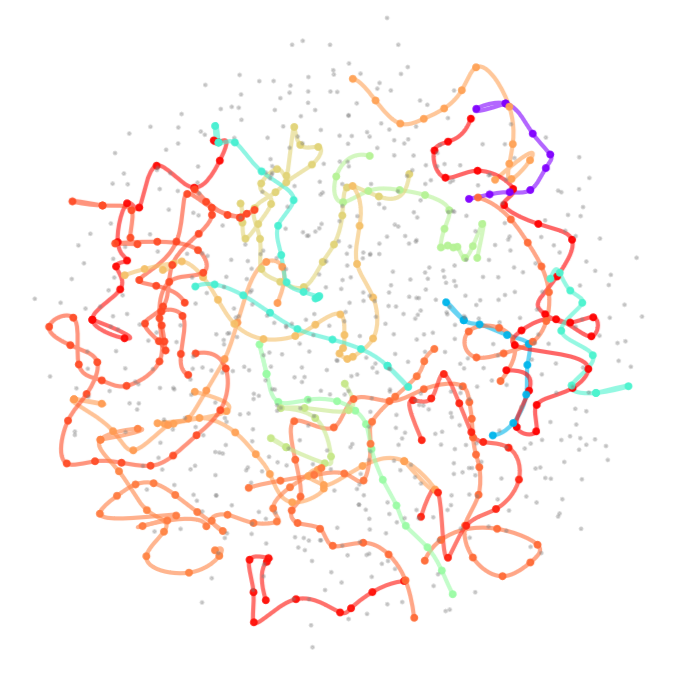}\caption{A CC of 1000 equal-mass particles in three dimensions with 24 filaments shown in different colours.}
\end{figure}

The fact that the filaments are structures which admit unambiguous identification and have different countable numbers of particles allows rigorous testing of the cosmological principle in CCs of sufficiently many particles in the range of ages in which filaments exist. For this we define the {\it sub-structure statistics} in any given balloon by means of the functions $n(l), ~V, ~I$, where $n$ is the number of filaments consisting of $l$ particles in the balloon, $V$ is the variety of the points within it, and $I$ their ingredient distribution (ratios of the particle species). Now suppose an infinite space containing infinitely many particles. We can say that in it the CP is satisfied to a prescribed accuracy if wherever a balloon is placed the values of $n(l), ~V, ~I$ differ from the mean found for a very large number of balloons by small prescribed percentages. Ongoing work by our collaborator Aaron Lax is exploring such a programme with CCs of up to 50, 000 particles and balloons containing 1000 particles. Of course, there is no guarantee that a distribution which passes the proposed test of its sub-structure statistics is exactly a CC but we think it likely will be close to one since CCs have critical values of the variety and, as Fig.~5 and the explanation of it shows, shapes that are nearly CCs occupy a substantial volume of the shape space to which they belong.

Before we continue, we need to define one more number to take into account the fact that it is not the number \no of particles that determines the variety $V$ but their $N(N-1)/2-1=M$ relative separations. This suggests definition of the {\it specific variety} as 
\be
\bar V:=V/M\label{spec}
\ee
with corresponding definition of specific age $\bar a$:
\be
\bar a=\bar V-\bar V_0.\label{ba}
\ee

With these definitions, this is our Law of the Universe:\footnote{It was anticipated at least in part in \cite{BarbourEtAl2024} on the basis of discussion between JB and Tim Koslowski.}

\vspace{.15in}

{\it It has infinitely many specific ages, \(\bar a\), with \(0\leq \bar a<\infty\). At each specific age, it contains infinitely many particles. These are distributed in such a way that the cosmological principle is qualitatively satisfied within regions (or `balloons') containing \(\mathcal C\) particles, where \(\mathcal C\) depends on the specific age. The proportions of the different ingredients within these regions are part of the law and are to be deduced from observation.}

\vspace{.15in}

If we discount the increase of the two-particle separations from the centre to the rim in the two CCs of 5000 thousand equal-mass particles, Fig. 9, (found by ML with a cluster at the Instituto de Astrofísica e Ciências do Espaço), they illustrate as clearly as one might wish at least at the `look-alike' level the fulfilment of our proposed law of the universe. In the top one, with variety extremely close to the absolute minimum, a disc containing $\mathcal C$ particles (with diameter say 1/10 of the CC) will cover look-alike regions wherever it is placed. For the bottom one, $\mathcal C$ must clearly be greater and the diameter of the disc containing $\mathcal C$ particles will be perhaps 1/5 of the CC's diameter. A similar effect is equally obvious in the 1000-particle CCs in Fig.~1. 

\begin{figure}[h!]
    \centering
    \includegraphics[width=.6\textwidth]{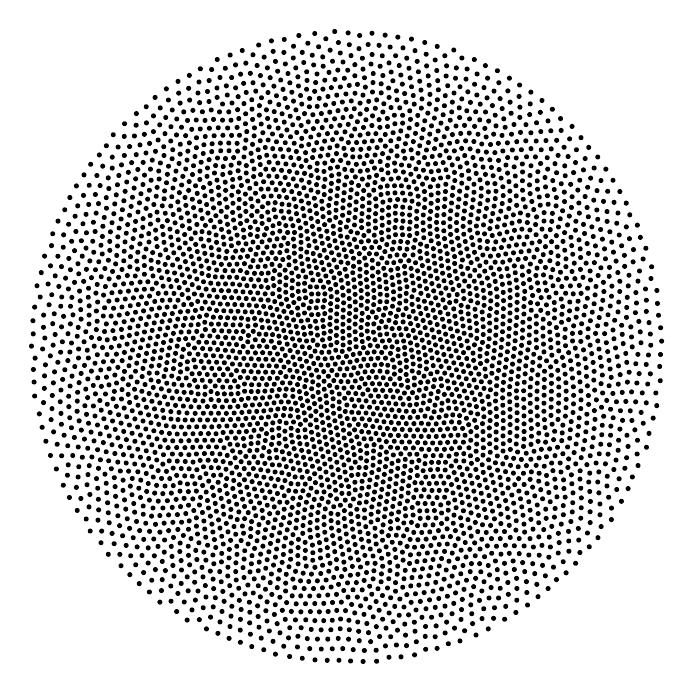}
    \vspace{3mm}
    \includegraphics[width=.6\textwidth]{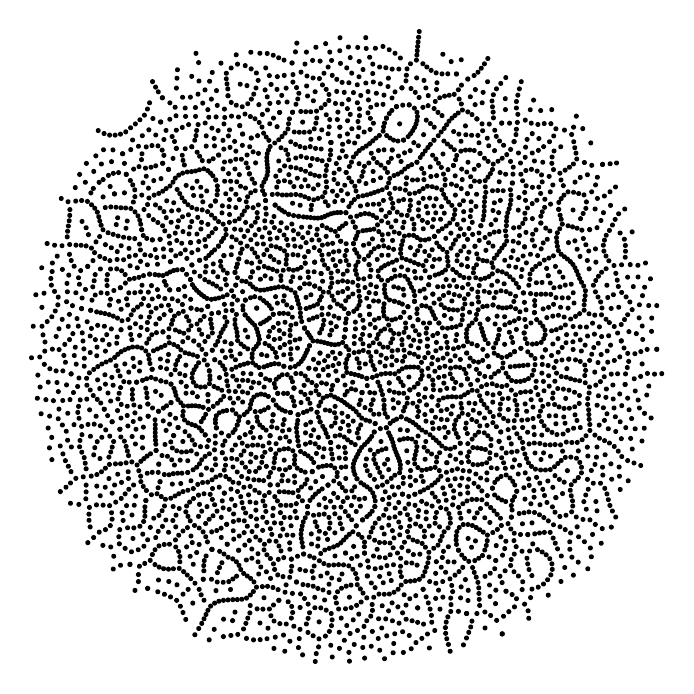}
    \caption{Two CCs of 5000 particles of equal mass. The one on the left is at or very near the minimum of the variety, the one on the right is about 1.7\% above it.}
\end{figure}

Before we continue we should admit a `sin': the finitely many particles of the CCs we have shown or indeed of one with however many particles, is not the infinite number of the Law of the Universe as formulated. In fact, as far as testing our ideas through computer simulations is concerned, at the time of writing we see no alternative to finding CCs with ever increasing numbers of particles and progressively increasing ages.

We consider it particularly significant that the CP, as we have defined it, appears to hold for a function that generalises the variety (\ref{fullb}) precisely in three-dimensional \eu space, and in no other dimension. This is a key defining feature of the simplest conceivable ontology with which we hope to explain the structure of the universe.

In support of this claim, we note first that, for uniform matter distributions, the shell theorem holds in two-dimensional \eu space for a potential with logarithmic distance dependence, with $1/r$ dependence in three dimensions, and $1/r^{-(n-2)}$ dependence in $n$-dimensional space for $n\ge 3$. However, we want to consider something rather different, namely the nature of the distribution of particles in dimensions 1, 2 and 3 in CCs obtained in all cases with a scale-invariant potential that, as so far considered, is a product of two factors with dimensions $[L]^1$ and $[L]^{-1}$, respectively, as is the case with the variety $V$ (\ref{fullb}). 

What we find especially interesting is that, at least for $V$ in 1, 2, and 3 dimensions fulfilment of the CP as we have defined it holds only in 3 dimensions. In dimension 1, Iszuierdo's collinear CC confirmed Lindstrom's estimate \cite{Lindstrom1998}, according to which the particle distribution has a finite length and density that is maximal in the centre and decreases symmetrically to zero at the two ends. The typical distributions that he found in two dimensions are shown in Fig.~1 and, as noted, have a particle density that decreases with distance from the centre. In contrast, and very strikingly, the particle density in three dimensions is, as shown in Fig.~7, uniform throughout a sphere with abrupt spherical edge. As we have defined it, the CP clearly holds for the equal-mass case in three dimensions within balloons containing sufficiently many particles. In the absence of numerical calculations, we can only speculate what might be the case in four or more dimensions, but as we live in three and our concern is with the simplest possible realistic point-particle ontology we will not discuss what might be the case in higher dimensions. We are also not yet in a position in which we have confirmed fulfilment of the CP for different particle ingredients but see no reason why it should fail.\footnote{We mention here how scale invariance changes one of the arguments advanced for heat death in modern cosmology based on the notion of an ever expanding universe, which, it is argued, will decrease the particle density forever leaving effectively nothing of interest. However, this conclusion relies on what we have called the scale fallacy. The variety, and with it the interest, of a system of point particles is scale invariant, as indeed is the meaning of Chinese characters and messages expressed through letters.}

This completes the account of the kind of law that we conjecture governs the universe that we currently observe and manifestly exhibits great variety but of specific kinds that, while vastly richer, can in principle be as readily identified as the tetrahedral shapes and filaments of the two ages that numerical work has so far enabled us to analyse and present. Besides that greater variety, which does not present any need to change in the principles of our variety-based notions, we also need to change at least somewhat the underlying theory from its \n origin to one that takes into account both of Einstein's relativity theories, in which, of course, the speed of light plays a decisive role. We say something about that in Sec.~4 of this paper and more in our companion paper \cite{BarbourLourenco}.

\textcolor{black}{We will conclude this part of the discussion by pointing out the immensity of the task our theory faces, while noting that it nevertheless seems capable, at least in principle, of undertaking it successfully. To appreciate both the scale of the challenge and the possibility of meeting it, we need only identify a few of the shape-statistics tasks that must be addressed and resolved.}

\textcolor{black}{Starting almost at random, we can consider galaxies of different kinds, containing different numbers of single, binary, and triple stars. These differences give rise to distinct statistics describing their substructure. Defining these statistics conceptually does not seem too difficult; demonstrating that a theory along the lines we have proposed can account for them would be a significant achievement. Yet this would constitute only a tiny part of what needs to be done, even if the overall task still seems, at least in principle, achievable.}

The law of this universe is quite unlike current theories. We will now argue, however, that it appears capable of being applied directly to the universe we currently observe, provided it can be modified along the lines outlined in the next section to reproduce the confirmed predictions of general relativity, rather than the results suggested by our analysis of Newtonian gravity. In our terminology, the observed universe clearly has an age \(a(s)\), as defined by (11), that is much greater than the ages of the CCs we have been able to construct numerically to illustrate our ideas.

In the next section we will draw attention to possible problems with it, both ones widely recognised and others suggested by the discussion so far presented in this paper. All we will say at this stage is that a law of the universe which is holistic is clearly going to be very different from one that is reductionist. In particular, one that is teleological and predicts ever increasing variety is by no means to be ruled out on {\it a priori} grounds. As we already noted and is spelled out in more detail in \cite{Barbour2020}, the rather unpleasant end predicted for the universe may simply reflect failure to consider the conditions under which the second law of thermodynamics can be assumed to hold.

\section{Absolutes in General Relativity} Much more could be said about Newton's absolutes, and before we turn briefly to the situation in Einsteinian gravity, we make one last comment about them. It is that, besides hugely reducing the predictive strength of the second law of motion, they may have given a quite wrong impression of what reality in our universe is like. One sees this already in the etymology of `real', which derives from the Latin \e{res} for `thing', and a thing can only be recognised by its attributes. Absolute space and time, being invisible, have none, or at least none that can be registered. Newton's reductionism made that impossible. The same defect may be lurking in the foundations of general relativity. Although its fundamental concept, space-time, does have variety expressed through curvature, it `disappears' in infinitesimal regions where it must become Minkowskian. This means that space-time, unlike a chain-link fence, is held together by nothing with real structure. Indeed, as Einstein commented, Minkowski space is as absolute as Newton's space and time are separately. The defining concepts of general relativity, proper distance and proper time, could be said to be directly lifted from Newton's absolute space and time, even if only in infinitesimal regions.

Conceptually, apart from the role ascribed to light (which we question in \cite{BarbourLourenco}), ontology had not changed where it counts. One sees the problem in Minkowski's famous lecture of 1908, in which he said ``The objects of our perception invariably include places and times in combination. Nobody has ever noticed a place except at a time, or a time except at a place'' \cite{LorentzEtAl1923}. Here there are already problems. How are places and times noticed, or, more fundamentally, how are they defined? Moreover, do places and times, as Minkowski clearly implies, have equal ontological status? That is questionable. To tell the time, we look to see where the hand of a clock is on the clock face. More fundamentally, any objective notion of place---of position---is relative. We need to be able to see things to know where we are; writing in the same period as Minkowski, Poincar\'e put great emphasis on that \cite{Poincare1902, Poincare1908}. Minkowski, for his part, was able, in the passage immediately following the words we have quoted, to avoid those problems because he was standing in front of a blackboard that served as a `materialisation' of invisible space and time as he said ``A point of space at a point of time, that is, a system of values \e{x, y, z, t,} I will call a \e{world-point.} The multiplicity of all thinkable \e{x, y, z, t} systems of values we will christen the \e{world}. With this most valiant piece of chalk I might project upon the blackboard four world-axes''. All very confident. However, Minkowski does exhibit some unease about his scheme because he says: ``Not to leave a yawning void anywhere, we will imagine that everywhere and everywhen there is something perceptible''. Just like Newton, the supreme reductionist, he has washed away all of reality and has had to sprinkle what we might call gold dust over his invisible space-time so that places can be identified. He shows no awareness of Mach's or Poincar\'e's critiques of Newton's concepts, both by then well known.

In creating general relativity, Einstein retained to great effect Minkowski's innovation of light cones but his introduction of variable space-time curvature only partially eliminated the need for Minkowski's gold dust since in infinitesimal regions space-time is Minkowskian. In those critical regions, it is the glue that, everywhere and everywhen, holds space-time together. However, since the glue is, apart from the light cones, featureless, one has to ask whether it can do the job. It may also be noted that many relativists are concerned that the field equations of general relativity allow an uncomfortably large number of solutions with many whose physical significance seems questionable. The analysis of solutions that the \nb allows suggests that `purification' of general relativity requires its ontology to be founded on recognisable things if the questionable solutions are to be eliminated. That could have far-reaching consequences.

In this connection, we note an interesting parallel with the additional degree of freedom $\kappa$ in the Newtonian $N$-body problem, discussed earlier, when the latter is freed from absolute structures and represented in shape space as a succession of shapes. The solution curves always bend away from any geodesic tangent to them, thereby tending towards ever higher values of the variety in the direction away from either a Janus point or a point of minimal \(V\). At such points, the solutions are maximally free of absolutes and, in Newtonian terms, may be of either the total-collision or the parabolic-escape type. The closest analogue of shape space in general relativity is conformal superspace. It was shown in \cite{BarbourOMurchadha2010} that something very similar happens in GR when its solutions are represented in conformal superspace: the solution curves are not geodesics but are subject to just one extra \(\kappa\)-type degree of freedom. This result was obtained shortly before the connection between such a non-geodesic degree of freedom and a non-entropic arrow of time in the \(N\)-body problem was recognised. Whether a similar connection can be established in GR has not been investigated, but seems entirely possible.

As the reader may be concerned that we implicitly retain a notion of simultaneity that most physicists think Einstein banished definitively, we note that Dirac \cite{Dirac1958}, in preparatory analysis undertaken in 1958 as the first step to quantisation of general relativity, felt bound to restore it on the basis of his Hamiltonian analysis of the dynamics that underlies general relativity. He pointed out that keeping the possibility of arbitrary slicing of space-time leads to the introduction of arbitrary degrees of freedom that are incompatible with quantum theory. At the end of his paper, which has attracted surprisingly little attention, he said that on the basis of his analysis he was inclined to believe that ``four-dimensional symmetry is not a fundamental property of the physical world''. In the light of Dirac's comment, we think our own suggestions need not appear too outlandish.

Although the bulk of our paper has been about gravity, the problem with reductionism and the boundary and initial conditions it enforces unnaturally on the universe can just as well apply to quantum mechanics \cite{BarbourLourenco}. Indeed, in it, obtaining physically meaningful solutions generally requires specifying appropriate boundary conditions. For example, the possible states of the hydrogen atom can only be obtained by requiring its wave function to vanish at spatial infinity.

Finally, reductionism goes entirely against the deep conviction of many thinkers that the universe is an undivided whole. Grasping such a totality directly is almost certainly beyond human capacity, and reductionism has successfully identified many of its parts. Perhaps it plays the role of Wittgenstein's ladder---enabling us to see, from the wall of the garden that is the universe, how its plants and borders fit together. But if Leibniz is right, we will never be able to grasp the infinite richness of the universe and discard the ladder. In the scheme we have sketched, the universe goes on getting evermore richly structured as its age, which has no upper bound, increases. In correspondence at the end of his life \cite{Leibniz1715}, Leibniz said he could not decide whether the universe is maximally varied at all times or strives to become evermore varied. In the scenario we have presented, there is no end to the increase of variety.

\section*{Acknowledgements}

MIRL gratefully acknowledges the Instituto de Astrofísica e Ciências do Espaço (IA) for providing access to the computing cluster used in this work. JB thanks Tim Koslowski for discussions and valuable input over many years and has been very lucky to have been enlightened about the finer details of $N$-body solutions by Richard Montgomery, Alain Chenciner, and Alain Albouy in particular, whose help in many discussions was decisive, above all in the finding of the remarkable central configuration of Fig. 1 by Manuel Izquierdo, whom we also thank for his contributions. We also thank Carmen Guard for support.

\end{document}